\documentclass[oneside,onecolumn,11pt]{extarticle}
\usepackage[utf8]{inputenc}
\usepackage[T1]{fontenc}
\usepackage[english]{babel}

\usepackage{extsizes}
\usepackage[left=2.5cm, right=2.5cm, top=1.785cm, bottom=2.0cm]{geometry}
\usepackage{setspace}
\usepackage[numbers,super,sort&compress,comma]{natbib}
\usepackage[usetitle]{rsc}
\usepackage[version=4]{mhchem}
\usepackage{siunitx}
\usepackage{mathptmx}
\usepackage{textcomp}
\usepackage{bm}
\usepackage{amssymb}
\usepackage{mathtools}
\newcommand{\degree}{^\circ}

\usepackage{graphicx}
\usepackage{float}
\usepackage{adjustbox}
\usepackage{rotating}
\usepackage{lscape}
\usepackage[format=plain,justification=justified,singlelinecheck=false,labelfont=bf,labelsep=space,font=doublespacing]{caption}

\usepackage{booktabs}
\usepackage{tabularx}
\usepackage{array}
\usepackage{multirow}
\usepackage{makecell}
\usepackage{threeparttable}
\usepackage{arydshln}

\renewcommand{\arraystretch}{1.15}
\newcolumntype{C}[1]{>{\centering\arraybackslash}p{#1}}
\newcolumntype{M}[1]{>{\centering\arraybackslash}p{#1}}
\newcolumntype{T}[1]{>{\centering\arraybackslash}p{#1}}
\newcolumntype{Y}{>{\centering\arraybackslash}X}

\usepackage{xcolor}
\usepackage{soul}
\definecolor{redhighlight}{RGB}{255, 200, 200}
\sethlcolor{yellow}

\usepackage{fancyhdr}
\renewcommand{\footnoterule}{%
  \kern -3pt
  \hrule width \textwidth height 1pt
  \kern 2pt
}
\renewcommand{\footnotesize}{\fontsize{10pt}{10pt}\selectfont}

\usepackage{enumitem}

\usepackage{xr-hyper}

\usepackage{hyperref}

\usepackage{ulem} 
\begin{document}

\vspace{4cm}

\noindent{\LARGE\textbf{Enhancing Conformality in Atomic Layer Deposition through Low Growth Per Cycle$\dag$}\par}


\noindent\large{
Christine Gonsalves\textit{*$^{a}$}, Jorge A. Velasco\textit{$^{a}$}, Ahmed Othman\textit{$^{b}$}, Ville Miikkulainen\textit{$^{b}$}, and  Riikka L. Puurunen\textit{*$^{a}$}


\let\thefootnote\relax\footnote{\textit{$^{a}$~Department of Chemical and Metallurgical Engineering, Aalto University,
P.O. Box 16100, FI-00076 AALTO, Finland.}}
\let\thefootnote\relax\footnote{\textit{$^{b}$~Department of Chemistry and Materials Science, Aalto University,
P.O. Box 16100, FI-00076 AALTO, Finland.}}

\footnotetext{$\dag$ Electronic supplementary information (ESI) available.}
\footnotetext{* Corresponding authors. E-mail: christine.gonsalves@aalto.fi, riikka.puurunen@aalto.fi}

\vspace{1em}

\clearpage

\noindent \textbf{Abstract} 

   Atomic layer deposition (ALD) is a key enabling technology for advanced microelectronics as it enables the growth of functional thin films on complex three-dimensional substrates with unmatched atomic-scale precision. Conformal film growth in high-aspect-ratio (HAR) structures is limited by slow diffusion of reactant molecules into deep, narrow features. This work elucidates an approach to grow conformal films faster by investigating the relationship between the ALD growth per cycle (GPC) and the film penetration depth in HAR structures through modeling and experiments. Diffusion-reaction simulations reveal, under Knudsen diffusion conditions, an inverse square root relationship between the film penetration depth and GPC. The prediction is validated experimentally with a model zinc oxide ALD process on rectangular lateral HAR structures, using an inhibitor molecule to decrease the GPC. While ALD is traditionally optimized for "high GPC," this work shows that "low GPC" may increase efficiency when conformality is key.

\section{Introduction}

 Atomic layer deposition (ALD) is an indispensable technique in advanced semiconductor manufacturing because it enables sub-nanometer thickness control and unparalleled conformality\cite{van_ommen_atomic_2021,puurunen_surface_2005,kessels2025atomic}. Excellent thickness control can be achieved, because the process relies on self-terminating reaction steps.\cite{van_ommen_atomic_2021,puurunen_surface_2005,kessels2025atomic} Conformality, meaning uniform film growth on all surfaces (especially challenging for non-line-of-sight regions and the very bottom of deep high-aspect-ratio (HAR) features) is critical for device functionality. Achieving conformal film growth especially on demanding HAR structures with aspect ratios exceeding 100:1, and complex three-dimensional geometries, depends heavily on the choice of process parameters.\cite{cremers2019conformality,yim2020saturation,yim2022conformality,gonsalves2024simulated,reiter2024modeling}  
Because mass transport inside deep HAR features proceeds via diffusion and the process is typically diffusion-limited~\cite{cremers2019conformality,gonsalves2024simulated}, a propagating saturation front forms inside the feature before conformality is achieved.

The growth per cycle (GPC) in ALD is characteristic of the process, and quantifies the amount of material added onto the surface in a single ALD reaction cycle.\cite{van_ommen_atomic_2021,puurunen2003growth}  Ideally, GPC is a function of the chosen reactants and the process temperature, and it may be initially affected by the substrate \cite{puurunen_surface_2005,puurunen2003growth,puurunen2003growththeot,kessels2025atomic,sonsteby2020consistency,richey2020understanding}. The GPC is most typically given as a thickness increment per cycle of the ALD-grown material MZ$_x$, $gpc_ {\mathrm{sat}}$ (nm), or increase in the  areal number density of atoms M, $q_0$ (\#/nm$^2$). The latter  can also be interpreted as the adsorption capacity of the surface  towards M in this particular ALD process.
The $gpc_ {\mathrm{sat}}$ and $q_0 $ are related:\cite{puurunen_surface_2005,ylilammi2018modeling,gonsalves2024simulated,yim2022conformality}
\begin{equation}
  q_0 = \frac{\rho\, gpc_{\mathrm{sat}} N_0}{M},
  \label{eq:q0_gpc}
\end{equation}
where $\rho\;(\mathrm{kg}\,\mathrm{m}^{-3})$ is the mass density of the ALD film material MZ$_x$ (M is metal, Z is nonmetal, $x$ is stochiometry), $N_0\;(\mathrm{mol}^{-1})$ is Avogadro's constant, and $M\;(\mathrm{kg}\,\mathrm{mol}^{-1})$ is the molar mass of one formula unit of the ALD-grown film material ($\mathrm{MZ}_x$). The $q_0$ is further inversely proportional to the averaged adsorption site area 
(containing a single payload atom M) on the substrate $a$; $q_0 = 1/a$. This relationship is illustrated in Figure \ref{fig:introscheme}a-d: when the adsorption capacity is halved, the average surface area occupied per M doubles (Figure \ref{fig:introscheme}b). A useful comparison point for the GPC is an average monolayer of the MZ$_x$ material, represented by cubes of MZ$_x$ units, with height equal to monolayer thickness $\bar{h}_{\mathrm{ml}}$ (Figure S1), base area $\bar{a}_{\mathrm{MZ}_x}$ ($\bar{a}_{\mathrm{MZ}_x} = \bar{h}_{\mathrm{ml}}^2$) and surface density $\bar{q}_{\mathrm{ml}}$ (M atoms in a monolayer of material MZ$_x$; $\bar{q}_{\mathrm{ml}} = \bar{h}_{\mathrm{ml}}^{-2}$). According to  recent reviews, the GPC in ALD is typically   up to about seven metal atoms per square nanometer \cite{piechulla2026atomic} or within 0.05-0.15 nm, \cite{kessels2025atomic}  corresponding to a fraction of a monolayer of the ALD-grown material \cite{puurunen2003growth, puurunen_surface_2005, kessels2025atomic}. Typically, a higher GPC is seen to be beneficial for process efficiency because more material added per cycle means less time needed to reach the target film thickness, resulting in improved processing speed. Much of ALD research, therefore has focused on strategies to \textit{increase} the ALD GPC. \cite{ham2022investigation,zhang2019high,tanaka2025growth,janocha2011ald}

\begin{figure}
    \centering
    \includegraphics[width=0.75\linewidth]{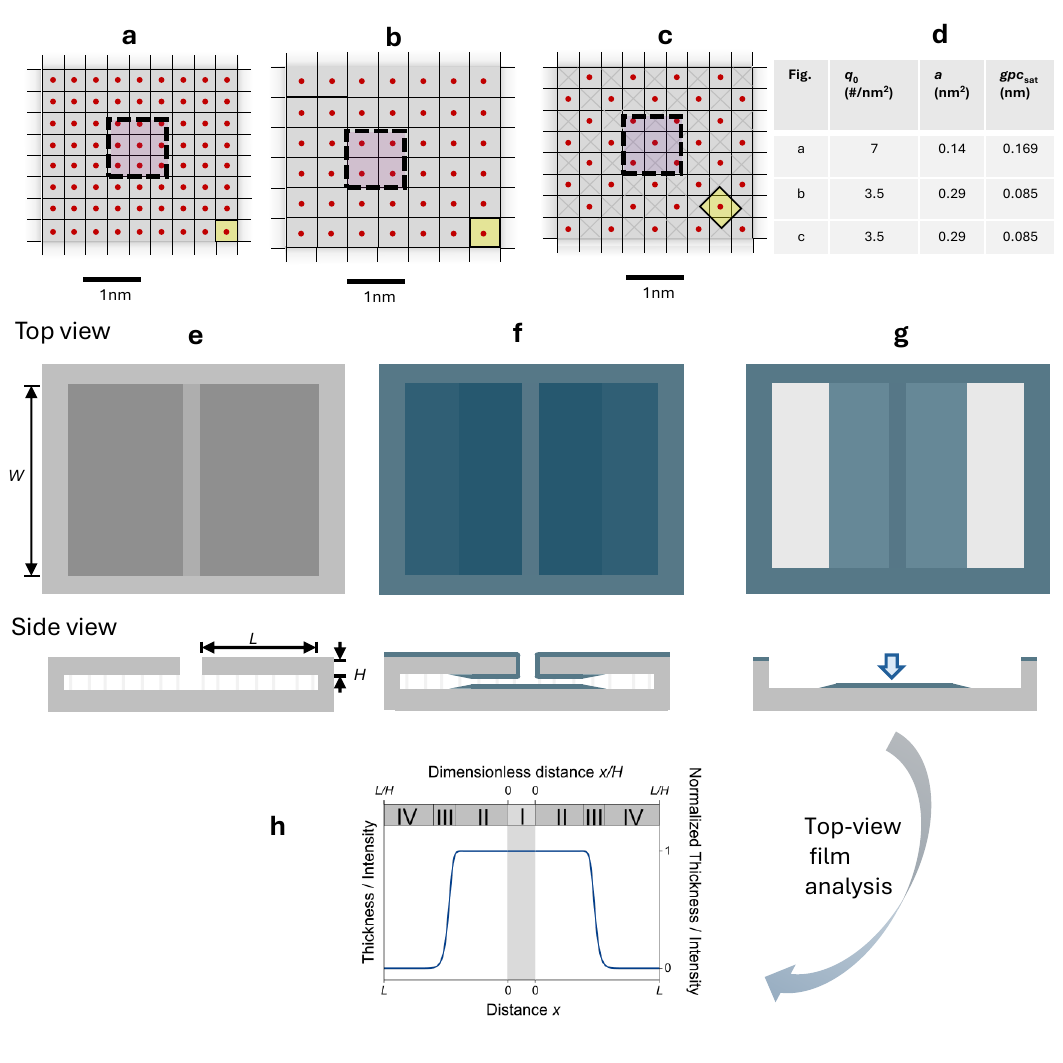}
    \caption{\textbf{Surface adsorption characteristics illustration, LHAR test structure schematics, and saturation profile zone classification.} (a) Schematic of the surface, where adsorption sites form a checker-box grid, where each site  can adsorb one metal M atom (to become material MZ$_x$ in a completed ALD reaction cycle). The black dashed square indicates a $1\text{ nm}^2$ reference box and small square with yellow shade indicates area $a$. This example has $a = 0.14\text{ nm}^2$ and corresponding adsorption capacity  $q_0 = 1/a = 7\text{ $\#$/nm}^2$. (b) Similar as panel (a), but with halved adsorption capacity ($q_0 = 3.5\text{ $\#$/nm}^2$) and double the adsorption site area. (c) Similar as panel (a), but half of the adsorption sites have been blocked by inhibitor (gray crosses), resulting in half of the GPC and double the adsorption site area. (d) Summary of the adsorption site area ($a$), adsorption capacity ($q_0$), and thickness-based GPC values ($gpc_{\mathrm{sat}}$) corresponding to panels (a)–(c). Values of $gpc_{\mathrm{sat}}$were calculated using Equation~\ref{eq:q0_gpc}, using a bulk density value ($\rho=$  5.61 g/cm$^3$) and molar mass ($M=$ 81.38 g/mol) corresponding to bulk ZnO~\cite{jensen2002x}. (e–g) Schematics of mirrored LHAR structures (height $H$, length $L$, and width $W$) shown: (e) before ALD, (f) after ALD, and (g) after peeling of the top membrane to expose the saturation profile for analysis. (h) Saturation profile classified into zones (as per Ref.~\cite{yim2020saturation}): Zone~I (open area in front of the channel), Zone~II (starting at channel entrance with $x=0$, typically saturated, extending up to a knee point), Zone~III (adsorption front with decreasing film thickness, and with the slope reflecting adsorption kinetics), and Zone~IV (unsaturated region with near-zero film thickness).}
   
    \label{fig:introscheme}
\end{figure}

 Scattered studies across  different ALD processes ($\text{Al}_2\text{O}_3$, $\text{TiO}_2$, and $\text{IrO}_2$)\cite{puurunen2016influence,mattinen2016nucleation} and simulations\cite{yim2022conformality,gonsalves2024simulated} have occasionally linked changing process parameters that are associated to decrease in the GPC, to an increased film penetration depth (Table S1 and S2). In those studies, the increasing film penetration typically arose as a secondary outcome of changing temperature, purge times, or type of reactants. The details of the correlation between GPC and film penetration depth have remained poorly described and understood, and are the focus of this study. To verify whether GPC directly affects film penetration depth, GPC must be varied independently and systematically.
 In simulations, GPC can be easily varied but intentionally varying GPC in experiments requires careful planning and modification of the process. A promising approach to decrease GPC is the introduction of inhibitor molecules,  such as alcohols, organic acids and $\beta$-diketones \cite{yu2024blocking,li2025competitive,yarbrough2021next,merkx2022relation,kytokivi1997reaction} to partially block adsorption sites (Figure \ref{fig:introscheme}c). Beyond their established role in area-selective ALD, inhibitors\cite{lee2025molecular,shearer2024role,mameli2023selection,ismaeel2025area} have also been shown to successfully decrease the GPC of many ALD processes  \cite{yanguas2013modulation,li2025competitive,merkx2022relation,lodha2024area,tezsevin2023computational,shearer2024role}.

In this work, we investigate in detail the relationship between GPC the film penetration in ALD both with simulations and experiments. Diffusion--reaction simulations are made for a large range of Knudsen numbers, representing Knudsen diffusion (Kn $\gg$ 1) and molecular diffusion (Kn $\ll$ 1) regimes. We selected the well-studied diethyl-zinc and water\cite{miikkulainen2013crystallinity} as a model ALD process to grow zinc oxide and reduced GPC in by introducing methanol as a site-blocking inhibitor molecule. Simulations predict an inverse square root relationship between ALD GPC and film penetration depth in the Knudsen diffusion regime. The prediction is validated by experiments and literature comparison.

\newpage

\section{Results}

Conformality analysis is performed by studying  ALD growth into rectangular lateral high-aspect-ratio (LHAR) structures that have such demanding aspect ratios, that the film does not penetrate to the end. This  results in \textit{partial conformality} and the formation of a \textit{saturation profile}. Simulations are made with a diffusion-reaction model and experiments with  PillarHall\textsuperscript{TM} test structures (often used for conformality tests\cite{cremers2019conformality,yim2020saturation,haimi2026atomic,philip2024conformal,van2023excellent,choolakkal2025using}), where the top membrane is removed to expose the thickness profile for detailed analysis (Figure~\ref{fig:introscheme}g). The characteristic thickness profile (often termed the saturation profile, Figure~\ref{fig:introscheme}h) can be represented in multiple forms (Section S2). Following classification from the literature \cite{yim2020saturation,yim2022conformality}, here, we present both \textit{as-measured profiles} (distance $x$ versus measured thickness or intensity) and \textit{Type I normalized} profiles (dimensionless distance $x/H$ versus normalized signal).

\newpage
\subsection{Simulation-based partial conformality studies}

Diffusion–reaction simulations were performed to analyze how varying the GPC influences film penetration depth. We implemented a one-dimensional continuum model describing gas transport into high-aspect-ratio structures based on Fick's law\cite{velasco2026ald}. The surface chemistry assumes a single-site Langmuir adsorption mechanism with reversible reactions where each reactive site binds one reactant molecule (see Section \ref{sec:methods} for full model description and Knudsen number classification). In this study, the adsorption capacity $q_0$ (representing the GPC, see Eqn \ref{eq:q0_gpc}) was varied from 0.5 to 8 $\#$/nm$^2$. To provide physical context, this corresponds to about 4\% to 67\% of a theoretical monolayer density of 12 Zn $\#$/nm$^{2}$ calculated for bulk $\mathrm{ZnO}$ ($\rho = 5.61\ \mathrm{g/cm^3}$, $M = 81.38\ \mathrm{g/mol}$)\cite{jensen2002x}. The simulations were performed for different sticking coefficients ($c$ = 0.1, 0,01, 0.0001), and different diffusion regimes; the Kn number was varied by 11 orders of magnitude (10$^{-5}$ to 10$^{5}$).

The simulated thickness profiles show that decreasing the adsorption capacity $q_0$ results in decreased film thickness and increased film penetration, demonstrating a general inverse correlation between film penetration and adsorption capacity (Figure~\ref{fig:expsim}a). From these profiles (examples shown as Figure S4), the penetration depth values across different adsorption capacities were compared (Figure~\ref{fig:expsim}b), and power-law regression fitting was performed to determine their detailed mathematical relationship (representative fits are in Figure S8). For Knudsen diffusion conditions (Kn $\gg1$, blue curves in Figure \ref{fig:expsim}b), there is an inverse square root relationship between the penetration depth and the adsorption capacity; i.e.,  $(x/H)_{\theta=0.5}$ = $C\,q_{0}^{-0.5}$. The value of the constant $C$ is dependent on the process conditions: the exposure, temperature and the gas properties.
For molecular diffusion conditions (Kn $\ll$1, red curves in Figure \ref{fig:expsim}b), the trend becomes more flat, reflecting a reduced sensitivity to changes in $q_0$ compared to Knudsen diffusion Figure \ref{fig:expsim}c). 
Nevertheless, a general inverse correlation between film penetration depth and adsorption capacity persists across all diffusion regimes and sticking coefficients (Figure S9).

\begin{figure}
    \centering
    \includegraphics[width=1\linewidth]{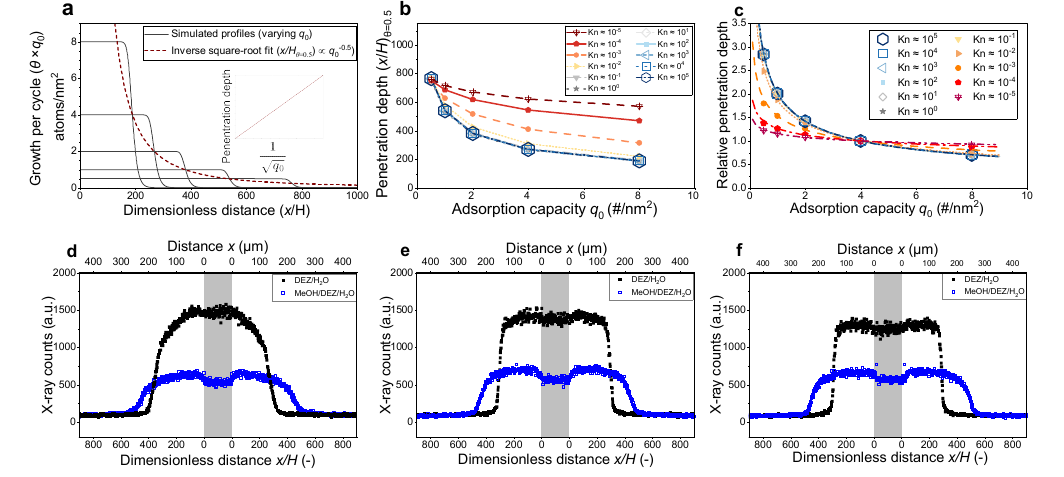}
    \caption{\textbf{Simulated and experimental partial conformality analysis results.} (a) Simulated saturation profiles in rectangular LHAR structures with varying adsorption capacity $q_0$ ($\mathrm{Kn} = 10^3$, $c = 0.01$). The dashed red line represents a power-law fit ($y = Cx^{-2}$, where $C = 1.433 \times 10^5$, $R^2 = 0.9999$, corresponding to $(x/H)_{\theta=0.5} \propto q_0^{-0.5}$) Other simulation parameters are shown in Table S4. (b) Penetration depth at half-coverage values, $(x/H)_{\theta=0.5}$, obtained from simulated saturation profiles across a range of Knudsen numbers ($c = 0.01$, exposures are shown in Figure S6a). (c) Relative penetration depth versus adsorption capacity for a broad range of Knudsen numbers. Penetration depth has been referenced to $q_0 = 4\text{ $\#$/nm}^2$. The lines show generalized power-law fits ($y = Cx^b$, $c = 0.01$, fit details are shown in Figure S6b). (d–f) As-measured SEM-EDS zinc saturation profiles for the mirrored rectangular LHAR structures after 180 ALD ZnO cycles at (d) 150 °C, (e) 175 °C, and (f) 200 °C. The black and blue curves denote the standard DEZ/water and methanol-inhibited ALD processes, respectively. The gray shaded region indicates the open area (Zone I), with zero marking the channel entrance (beginning of Zone II). The corresponding Type-I normalized profiles are shown in Figure S10.}
    
    \label{fig:expsim}
\end{figure}

 \subsection{Experimental partial conformality studies}


To complement the simulations with experiments, a series of ALD runs was designed where the GPC was intentionally decreased using an inhibitor molecule and the penetration depth in LHAR structures was analyzed. The widely studied DEZ/water-based process to grow zinc oxide\cite{sadruddin2026designing, tynell2014atomic}, for which the first saturation profiles have already been published\cite{haimi2026atomic},  was chosen for testing. Initially, methanol and acetylacetone (Hacac) were tested as inhibitors molecules. Hacac resulted in practically no growth in accord with literature \cite{mameli2023selection}, while methanol decreased the GPC, and methanol was selected for further investigations. In a typical ZnO process, DEZ reacts with the surface hydroxyl groups, loses ethane, and results in an ethyl-terminated  surface\cite{weckman2018atomic,cai2019revisit}. In the methanol-inhibited case, we expect some hydroxyl sites to react with methanol to leave behind a methoxy group and release water. This will lower the ZnO GPC defined by the DEZ step, as hydroxyls are already consumed and cannot react with DEZ. In the second ALD step, water is expected to react with both ethyl and methoxy groups, to leave behind hydroxyl groups (and release ethane and methanol). In the experiments, care was taken to ensure that the DEZ and water exposures (partial pressure $\times$ time) remained constant by appropriate adjustment of the
purge times (see Section \ref{sec:ALD_expts} for details).

According to measurements made on planar wafers, methanol worked successfully as an inhibitor in the DEZ/water-based  ALD ZnO process.  The GPC of the standard non-inhibited process was on the order of 0.15 nm (well in line with the literature \cite{tynell2014atomic, haimi2026atomic, kim2011properties, jeon2008structural,sadruddin2026designing}) and decreased slightly when temperature was increased from 150 to 200~$^{\circ}$C. The GPC was approximately halved by the use of methanol as inhibitor at all tested temperatures (ellipsometry results in Table \ref{tab:GPC_PD} and XRR results Table S11).

\begin{sidewaystable}[p]
\centering
\begin{threeparttable}
\caption{Thickness, GPC, penetration depth, film density, and adsorption metrics for (uninhibited) DEZ/H$_2$O and (inhibited) MeOH/DEZ/H$_2$O processes. All experiments correspond to 180 ALD cycles.\protect\tnote{a}}
\label{tab:GPC_PD}

\setlength{\extrarowheight}{4pt} 
\renewcommand{\arraystretch}{1.3}
\setlength{\tabcolsep}{4pt} 
\footnotesize

\begin{tabular}{|c|cc|cc|cc|cc|cc|c|c|}
\hline
\multirow{2}{*}{\makecell[c]{ALD\\temp.\\($^\circ$C)}} &
\multicolumn{2}{c|}{\makecell[c]{Thickness from\\ellipsometry (nm)}} &
\multicolumn{2}{c|}{\makecell[c]{GPC from\\ellipsometry (nm/cycle)}} &
\multicolumn{2}{c|}{\makecell[c]{Film density from\\XRR (g/cm$^3$)}} &
\multicolumn{2}{c|}{\makecell[c]{Penetration depth from\\SEM-EDS\tnote{b} $(x/H)_{\theta=0.5}$ (-)}} &
\multicolumn{2}{c|}{\makecell[c]{Adsorption capacity\\$q_0$ in Zone IIb (\#/nm$^{2}$)}} &
\multirow{2}{*}{\makecell[c]{Decrease in\\adsorption capacity\\$q_0$ in Zone IIb\\with inhibitor (\%)}} &
\multirow{2}{*}{\makecell[c]{Increase in\\penetration depth\\with inhibitor (\%)}} \\
\cline{2-11} 
 &
\makecell[c]{DEZ/\\[-2pt]water} & \makecell[c]{MeOH/DEZ/\\[-2pt]water} &
\makecell[c]{DEZ/\\[-2pt]water} & \makecell[c]{MeOH/DEZ/\\[-2pt]water} &
\makecell[c]{DEZ/\\[-2pt]water} & \makecell[c]{MeOH/DEZ/\\[-2pt]water} &
\makecell[c]{DEZ/\\[-2pt]water} & \makecell[c]{MeOH/DEZ/\\[-2pt]water} &
\makecell[c]{DEZ/\\[-2pt]water} & \makecell[c]{MeOH/DEZ/\\[-2pt]water} &
 &  \\
\hline
150 & 29.7 & 12.9 & 0.165 & 0.072 & 5.3 & 5.3 & 297 $\pm$ 4 & 430 $\pm$ 3 & 6.6 & 3.3 & 49.2 & 44.6 \\
\hline
175 & 28.4 & 12.4 & 0.158 & 0.069 & 5.2 & 5.3 & 298 $\pm$ 2 & 428 $\pm$ 2 & 6.2 & 3.3 & 46.7 & 43.9 \\
\hline
200 & 25.4 & 12.0 & 0.141 & 0.067 & 5.0 & 5.4 & 288 $\pm$ 1 & 454 $\pm$ 5 & 5.5 & 3.1 & 43.2 & 57.7 \\
\hline
\end{tabular}

\begin{tablenotes}[flushleft]\footnotesize
\item[a] Other measurement results (thickness, roughness) from XRR are in Table S11 of the Supporting Information.
\item[b] Average over six line scans (see Table S14 in the Supporting Information)
\item[c]  The $q_0$ was calculated by multiplying the $q$ with the ratio of the Zn signal intensity measured in the SEM-EDS in Zone IIb/Zone I of the saturation profile. The signal intensities are tabulated in Table S13. The value of the experimental adsorption capacity $q$ was calculated from the GPC measured by ellipsometery and the density (from XRR) using Equation \ref{eq:q0_gpc} ).
\end{tablenotes}
\end{threeparttable}
\end{sidewaystable}

Partial conformality analysis was performed for the ALD ZnO films with and without methanol inhibitor, at three temperatures, using   PillarHall$^\mathrm{TM}$ conformality test chips, placed in the reactor along with the flat wafers in the same run (details are in the Methods section). Figure \ref{fig:expsim}d,e,f shows representative SEM-EDS line scans for measured Zn intensity profiles after removal of the top membrane of the (mirrored) rectangular LHAR test structure, with a central open area Zone I (indicated with a gray background in Figure \ref{fig:expsim}d,e,f). Further details on the saturation profile analysis are presented in the supporting information (Section S4.6). 

Methanol inhibition decreased the GPC in ZnO ALD inside the LHAR channels, with the Zn signal intensity dropping to less than half, as seen in the as-measured intensity profiles Figure \ref{fig:expsim}d,e,f (detailed numerical values are in Table S13, Type-I normalized profiles are in Figure S10). In the methanol-inhibited ZnO process, the film penetrated roughly 50\% deeper than in the standard ZnO process (detailed numerical values in Table \ref{tab:GPC_PD}). The experimental penetration depth at half-coverage as a function of the adsorption capacity $q_0$ obtained from ellipsometry and XRR, is plotted alongside the simulated inverse square-root curve $\left((x/H)_{\theta=0.5} = C\,q_0^{-0.5}\right)$ and is shown as Figure S11. The experiments validate the trend predicted in the simulations.

\section{Discussion}

To benchmark the observed inverse square root relationship against earlier experimental and simulated conformality ALD data, Figure \ref{fig:compare} shows literature values alongside this work's experimental data and the predicted trend. All datasets are normalized with respect to a reference sample (sample 1, S1; Supporting Information, Table S2), making $(1.0, 1.0)$ the reference anchor. Moving towards the left along the horizontal axis, a decrease in the normalized GPC corresponds to a non-linear increase in the relative penetration depth. In general, the predicted trend  holds across all datasets. Some experimental data points (black circles) show some scatter above the theoretical curve, however, the general inverse correlation still holds. The experiments of DEZ/water ALD carried out in this work (red stars) are in line with the predicted inverse square root relationship and also with existing literature.

\begin{figure}[h]
    \centering
    \includegraphics[width=0.5\linewidth]{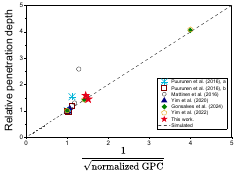} 
    \caption{Relative penetration depth ($x_{\theta=0.5} / x_{\theta=0.5, \text{ref}}$) as a function of normalized growth per cycle ($\text{GPC}/\text{GPC}_{\text{ref}}$). The inverse square root relationship $y = x^{-0.5}$ is shown as a gray dashed line ("simulated"). To compare the different works, each dataset was normalized against its own baseline sample (designated as sample 1, S1 in the supporting information, Table S2), where GPC was represented either by adsorption capacity ($q_0$) or thickness increment per cycle ($gpc_\text{sat}$). Comparison of ALD processes from literature: 
    TiO$_2$ (Puurunen et al., 2016a \cite{puurunen2016influence}), 
    Al$_2$O$_3$ (Puurunen et al., 2016b \cite{puurunen2016influence}), 
    metallic Ir (Mattinen et al., 2016 \cite{mattinen2016nucleation}),  
    Al$_2$O$_3$ (Yim et al., 2020 \cite{yim2020saturation}); 
    simulated works inspired from the trimethylaluminum--water ALD process 
   (Gonsalves et al., 2024 \cite{gonsalves2024simulated} and 
    Yim et al, 2022 \cite{yim2022conformality}). ZnO data points with and without inhibitor shown as "this work".  }
    \label{fig:compare}
\end{figure}

One may find the inverse correlation between GPC and penetration depth counterintuitive --- after all, "high GPC" has historically been sought for ALD. The experimental analysis presented in this work along with literature comparison validates the trend to be real and generic. How to understand this correlation, in simple terms? 

First, understanding the generic \textit{inverse correlation} is facilitated by considering ALD growth to progress through adsorption of species rather than as addition of thickness; i.e., consider GPC through the adsorption capacity $q_0$ rather than thickness increment $gpc_\mathrm{sat}$ (the two are directly proportional through the mass density,  Eqn. \ref{eq:q0_gpc}). Second, consider the average area $a$ that an adsorbing species containing  one payload atom M (Figure~\ref{fig:introscheme}a-d) will occupy; $a=1/q_0$ (the area is increased both by the size and number of ligands attached to the payload atom\cite{puurunen2003growth}, and also by inhibitors blocking some of the reactive sites, see Figure~\ref{fig:introscheme}c). An amount of molecules $n$ carrying the payload atom that enter the high-aspect-ratio feature  will cover an area $A$ that is equivalent to the area occupied by n molecules ($A=na$). As the feature width $W$ does not change, the coating depth $x_{\mathrm{PD}}$ will be determined by the area: $Wx_{\mathrm{PD}}=na$, yielding  $x_{\mathrm{PD}}=na/W$. Further replacing $a$ with $q_0$ in the equation, directly delivers the inverse correlation, $x_{\mathrm{PD}}=n/(Wq_0)$. A generic inverse correlation therefore follows directly from the area occupied by a unit of adsorbed species. 

Next, to understand the specific \textit{inverse square root relationship}, one must realize that the amount of reactant molecules entering the LHAR cavity is not constant with time; it depends on the penetration depth as described in Fick's law. The further the film propagates, the smaller the partial pressure gradient, that is the driving force for diffusion (partial pressure gradient is inversely proportional to the penetration depth, see e.g. Figure S5), and thereby the slower the diffusion. The inverse square root relationship of penetration depth on GPC follows from combining that (i) the  coating time (for a given penetration depth) is directly proportional to the adsorption capacity (i.e. GPC) and (ii) the coating time increases with the square of penetration depth into the LHAR cavity. 

The inverse square root relationship can be  derived from the analytical Gordon et al. model \cite{gordon2003kinetic} (and similarly from the Ylilammi et al. \cite{ylilammi2018modeling} model) for Knudsen diffusion conditions.  A full mathematical derivation of the inverse square root relationship from the analytical Gordon et al. \cite{gordon2003kinetic} is presented in SI (Section S5); a short version is presented here. Using the symbol notation as in this work and after interpreting the "aspect ratio" of Gordon et al. \cite{gordon2003kinetic} for the "infinitely wide" rectangular LHAR geometry of this work, we get from Equation 13 of Gordon et al.\cite{gordon2003kinetic} to:
\begin{equation}
    p_{A0} t = q_0 \sqrt{2\pi m k_\mathrm{B} T} \times \frac{3}{2} \left( \frac{x_{\mathrm{PD}}}{2H} \right)^2 .
\label{eq:gordon-rewritten}
\end{equation}

\noindent Solving  for $x_{\mathrm{PD}}$ (penetration depth, meters), we get:

\begin{equation}
    x_{\mathrm{PD}} = \sqrt{\frac{8}{3} \frac{ p_{A0} t}{q_0 \sqrt{2\pi m k_\mathrm{B} T}}} \cdot H .
\label{eq:inverse-square-root-derived}
\end{equation}
\noindent This  is exactly the sought-after inverse square root relationship, $x_{\mathrm{PD}} \propto q_0 ^{-0.5}$.

The benefit of a lower-GPC process can be illustrated by considering two processes where merely the adsorption capacity differs. Consider two cases with GPC values $q_{0,1}$ and $q_{0,2}$, where $q_{0,1} > q_{0,2} $. For any HAR structure where the above-presented equations are valid, the lower GPC reduces the exposure time required to saturate the structure: $ t \propto q_0$ from Eq.\ref{eq:gordon-rewritten}; $t_{2}/t_{1} = q_{0,2}/q_{0,1} $. Alternatively, the lower GPC  results in higher penetration depth for the same exposure: $x_{\mathrm{PD}} \propto \sqrt{1/q_0}$ from Eq.\ref{eq:inverse-square-root-derived}; $ x_{\mathrm{PD},2}/x_{\mathrm{PD},1} = \sqrt{(q_{0,1}/q_{0,2})} $. To give a realistic  numerical  example, we consider adsorption-capacity-based GPC of 7 and 2 \#/nm$^2$, which are typical for Zn in ZnO processes, as found e.g. in this work and literature \cite{yim2023atomic}. In this case, the lower GPC reduces the exposure time required to saturate the structure in a reaction step by 71\% ($t_{2}/t_{1} = q_{0,2}/q_{0,1} = 2/7 =0.286$), or results in an 87\% higher penetration depth for the same exposure 
($ x_{\mathrm{PD},2}/x_{\mathrm{PD},1} = \sqrt{(q_{0,1}/q_{0,2})} = \sqrt{(7/2)} =1.87 $).

\section{Conclusions and Outlook}

This work showed by  simulations,  experiments and comparison to literature that for a constant exposure (reactant's partial pressure $\times$ time) and when operating in the Knudsen diffusion regime under diffusion-limited conditions, the film penetration depth increases with a decreasing GPC through an inverse square root relationship.

Producing conformal films efficiently is necessary for current and future industrial applications of ALD. The GPC in ALD can be decreased and conformality thereby enhanced in (at least) three ways: (i) using inhibitor molecules to block a fraction of surface reactive sites and thereby lower the GPC (as done in this work); (ii) tailoring the metal reactant's size and reactivity to give a lower GPC; and (iii) modulating the ALD temperature to decrease GPC. The effect of temperature on GPC may be complex, however: temperature not only affects GPC (and adsorption kinetics) but also gas density and diffusion, and as shown previously by simulations, \cite{yim2022conformality, gonsalves2024simulated, heikkinen2024atomic} merely increasing temperature without affecting GPC should actually decrease the penetration depth.  

In the past, "high-GPC" ALD processes have most often been sought. In the future, when conformality on complex three-dimensional substrates will be in focus, "low-GPC" processes may be deliberately sought to achieve fast conformal filling, resulting in a lower carbon footprint \cite{weber2023assessing} and improved overall  sustainability.

\section{Acknowledgments}

The authors gratefully acknowledge Aaro Leppänen for part of the ZnO ALD experiments, Eetu Varttila for valuable discussions related to data processing, and Robin Ras for helpful suggestions on the manuscript. Okmetic Oy supplied the Si wafers  and Chipmetrics Oy provided the PillarHall$^{TM}$ LHAR chips for this study. For financial support, C.G. thanks the Vilho, Yrjö and Kalle Väisälä Foundation of the Finnish Academy of Science and Letters; A.O. and V.M. the Miikkulainen starting grant at Aalto University; C.G., J.A.V. and R.L.P. the GENESIS project under Grant Agreement No. 101194246. GENESIS is an EU program supported by CHIPS JU and its associated member states. In addition, GENESIS Swiss partners are supported by the Swiss state secretariat (SERI). Views and opinions expressed are however those of the author(s) only and do not necessarily reflect those of the European Union or Chips JU. Neither the European Union nor the granting authority can be held responsible for them. Computational resources were provided by the Aalto Science-IT services, and CSC – IT Center for Science, Finland.

\section{Conflicts of interest}
R.L.P. is inventor in a PillarHall\texttrademark-related patent.

\section{Data availability}
Experimental saturation profile data will be available openly on Zenodo, (DOI: to come here), included in the https://zenodo.org/communities/ald-saturation-profile-open-data/ community. The code used for the simulations is made openly available on GitHub\cite{velasco2026ald}.

\section{Author contributions}

The inverse square root relationship was jointly discovered by C.G. and R.L.P. The full diffusion-reaction simulation code used in this work was developed by J.A.V upon request by R.L.P. Together, C.G., J.A.V., and R.L.P. planned the simulations, which C.G. performed. C.G., R.L.P. and V.M. jointly planned the ALD experiments, which A.O. performed under supervision of V.M. Likewise, A.O. made XRR, and ellipsometry measurements, and V.M. supervised. C.G. performed the SEM-EDS measurements, conducted the saturation profile analysis, and compiled the first version of the manuscript under supervision by J.A.V. and R.L.P. All authors participated in interpretation of data and contributed to writing the manuscript.

\clearpage

\section{Methods}
\label{sec:methods}

\subsection{Description of the model}

In this work, we use a one-dimensional diffusion-reaction model to describe gas transport into a high-aspect-ratio structure\cite{velasco2026ald}. The model follows a single-site Langmuir adsorption model, where each reactive site binds to one reactant molecule, and assumes that the surface reactions are reversible. The transport of the reactant gas is modeled using a diffusion-reaction equation based on Fick's law with an additional adsorption loss term. Diffusion of the reactant gas molecule with pressure $p_A$ along a distance $x$ is described by\cite{gonsalves2024simulated,yim2022conformality,ylilammi2018modeling}:
\begin{equation}
    \frac{\partial p_{A}}{\partial t} = D_{\mathrm{eff}}\,\frac{\partial^{2} p_{A}}{\partial {x}^{2}} - \frac{4 g R T}{h N_{0}},
    \label{eq:pressure_diff}
\end{equation}

and the evolution of surface coverage is given as\cite{gonsalves2024simulated,yim2022conformality,ylilammi2018modeling}

\begin{equation}
    \frac{d\theta(x,t)}{dt} = \frac{1}{q_0}\, c Q p_A \,(1-\theta) - P_d \,\theta 
    \label{eq:surface_coverage}.
\end{equation}

In Equation \ref{eq:pressure_diff}, $D_{eff}$ is the effective diffusion coefficient (m$^2$s$^{-1}$); $g$ (m$^2$s$^{-1}$) is the net adsorption rate i.e., the difference between the rate of adsorption and desorption; $R$ (Pa $\cdot$ m$^3$ mol$^{-1}$ K$^{-1}$) is the gas constant; and \textit{T} (K) is the reaction temperature. The hydraulic diameter is given by  $h$ (m), which for an LHAR structure is calculated from the width $W$ (m) and channel height $H$ (m) using the relation $h = 2/((1/H)+(1/W))$. $Q$ is the collision rate at unit pressure ($\text{m}^{-2}\ \text{s}^{-1}\ \text{Pa}^{-1}$), given as\cite{yim2022conformality,gonsalves2024simulated,ylilammi2018modeling}
\begin{equation}
    Q = \frac{N_0}{\sqrt{2\pi M_{\text{A}}RT}},
\end{equation}
where $M_{\text{A}}$ ($\text{kg}\ \text{mol}^{-1}$) is the molar mass of reactant A.

The different diffusion regimes discussed, in this work, are characterized by the Kn number\cite{gonsalves2024simulated,cremers2019conformality,ylilammi2018modeling}. i.e., the ratio of the mean free path $\lambda$ to the characteristic limiting feature dimension \textit{D} as in \cite{gonsalves2024simulated,yim2022conformality,ylilammi2018modeling}:
\begin{equation}
   Kn = \lambda / D 
   \label{eq:Kn_number}
\end{equation}
When Kn $\gg$1, molecule-wall collisions dominate and transport is said to be in the Knudsen diffusion regime. And when Kn $\ll$1, molecule-molecule collisions dominate and transport takes place via. molecular diffusion, with the transition region Kn$\sim$1 between the two. The mean free path $\lambda$ of the reactant gas A in a system of two gases (reactant A and inert gas I) is\cite{gonsalves2024simulated,yim2022conformality,ylilammi2018modeling}
\begin{equation}
  \lambda
  = \frac{k_B T}{
      \sqrt{2}\, p_{A0}\, \sigma_{A,A}
      + \sqrt{1 + \frac{m_A}{m_I}}\, p_I\, \sigma_{A,I}
    }
  \label{eq:meanfree}
\end{equation}
 In Equation \ref{eq:meanfree},  $k_B$ (J K$^{-1}$) is the Boltzmann constant, $T$ (K) is the temperature, $p_{A0}$ and $p_I$ (Pa) are the partial pressures of reactant A and inert gas I respectively; $m_A$ and $m_I$ (kg) are the molecular masses of the reactant and inert gas respectively; and $\sigma_{A,A}$ and $\sigma_{A,I}$ are
the collision cross sections (m$^2$) between species indicated by the subscripts. The collision cross section between two molecules, say $i$ and $j$ is given as:\cite{gonsalves2024simulated,yim2022conformality,ylilammi2018modeling}
\begin{equation}
  \sigma_{i,j}
  = \pi \left( \frac{d_i}{2} + \frac{d_j}{2} \right)^2,
  \tag{5}
\end{equation}
in which $d_i$ and $d_j$ (m) are the hard-sphere diameters of molecules $i$ and $j$,
respectively.

The effective diffusion of a gas, the $D_{\mathrm{eff}}$ in Equation \ref{eq:pressure_diff} can be calculated from the Bosanquet relationship:\cite{gonsalves2024simulated,ylilammi2018modeling,yanguas2012self}

\begin{equation}
\frac{1}{D_{\mathrm{eff}}} = \frac{1}{D_{A}} + \frac{1}{D_{\mathrm{Kn}}}.
\label{eq:Deff}
\end{equation}
Here, $D_{\mathrm{Kn}}$ (m$^2$s$^{-1}$) is the Knudsen diffusion coefficient, which dominates in Knudsen diffusion conditions where the pressure is low and molecule-wall collisions dominate; given by\cite{gonsalves2024simulated,yim2022conformality,ylilammi2018modeling}
\begin{equation}
D_{\mathrm{Kn}} = h \sqrt{\frac{8 R T}{9 \pi M_{\mathrm{A}}}},
\label{eq:Dkn}
\end{equation}
where $M_{\mathrm{A}}$ is the molar mass, (kg\,mol$^{-1}$) of the reactant A. 
The $D_{\mathrm{A}}$ in Equation \ref{eq:Deff} is the molecular diffusion coefficient that describes gas phase collisions, and is given as\cite{gonsalves2024simulated,yim2022conformality,ylilammi2018modeling}

\begin{equation}
D_{\mathrm{A}} = \frac{3\pi\,\bar{v}_{\mathrm{A}}^{\,2}}{16\,z_{\mathrm{A}}}.
\label{eq:DA}
\end{equation}
The thermal velocity $\bar{v}_{\mathrm{A}}$ (m\,s$^{-1}$)  is given by\cite{gonsalves2024simulated,yim2022conformality,ylilammi2018modeling}
\begin{equation}
\bar{v}_{\mathrm{A}} = \sqrt{\frac{8RT}{\pi M_{\mathrm{A}}}},
\label{eq:vA}
\end{equation}
and the collision frequency $\tilde{z}_{\mathrm{A}}$ is
\begin{equation}
z_{\mathrm{A}} = \frac{\pi}{4}\,\bigl(d_{\mathrm{A}}+d_{\mathrm{I}}\bigr)^{2}
        \sqrt{\frac{8RT}{\pi}\!\left(\frac{1}{M_{\mathrm{A}}}+\frac{1}{M_{\mathrm{I}}}\right)}
        \frac{p_{\mathrm{I}} N_{0}}{RT}
        \;+\;
        \pi d_{\mathrm{A}}^{2}\,
        \sqrt{\frac{16RT}{\pi M_{\mathrm{A}}}}\,
        \frac{p_{\mathrm{A}} N_{0}}{RT},
\label{eq:zA}
\end{equation}
where $M_{\mathrm{I}}$ is the molar mass of the inert gas (kg\,mol$^{-1}$); $p_{\mathrm{I}}$ (Pa) is the partial pressure of the inert gas;
$d_{\mathrm{A}}$ and $d_{\mathrm{I}}$ (m) are the molecular diameters of the reactant and the inert gas, respectively.

\subsection{Simulation details}
The diffusion-reaction model equations (Eq. \ref{eq:pressure_diff} and \ref{eq:surface_coverage}) were simulated using a Python implementation\cite{velasco2026ald}. All simulations were performed for one ALD reactant pulse, assuming it is the limiting step and represents an ALD cycle. In the simulations, the adsorption capacity $q_0$  was varied from 0.5 to 8 \#/nm$^{2}$. From the resulting saturation profiles, penetration depth at half coverage ($(x/H)_{\theta =0.5}$) was estimated. To analyze the relation between penetration depth and adsorption capacity, a nonlinear power-law regression of the form $y = Cx^b$ was done using Microsoft Excel, where $y$ is $(x/H)_{\theta =0.5}$ and $x$ is the adsorption capacity $q_0$. Simulations were done at different diffusion regimes with Kn numbers from the free molecular flow regime (Kn number 2.37 $\times 10^{5} $) up to the continuum regime (Kn number 2.37 $\times 10^{-5} $). Simulations were done using three (lumped) sticking coefficients $c$ = 0.1, 0.01, and 0.001. To get to different Kn number regimes, we varied the channel heights (from 10$^{-3}$ to 10$^{-8}$ m) and exposures by varying both the partial pressure and time (from  2.6 $\times 10^{7} $ to 10 Pa $\cdot$ s). A summary of the detailed simulation parameters is presented in the supporting information (Table S4).

\subsection{Atomic layer deposition experiments}
\label{sec:ALD_expts}
Zinc oxide (ZnO) thin films were grown on Si wafers in a Picosun R-200 ALD reactor by atomic layer deposition (ALD) using diethylzinc (DEZ, Zn(C$_2$H$_5$)$_2$, $\geq$95\%, Strem Chemicals) and deionized water (H$_2$O) as reactants. N$_2$ (99.999\%, Linde) was used as the carrier gas. Both precursors were delivered at 22$^{\circ}\text{C}$. The ALD process temperatures were 150, 175, and 200 $^{\circ}\text{C}$, with corresponding reaction chamber pressures measured at $0.92 \pm 0.01$, $0.90 \pm 0.01$, and $0.86 \pm 0.01\text{ hPa}$, respectively. The pressure in the intermediate space (IMS) between the reaction chamber and the outer vacuum chamber was 14 hPa. The pulse and purge sequence for the DEZ/H$_2$O process was DEZ pulse–purge–H$_2$O pulse–purge, with pulse and purge times being 0.2 - 26 - 0.2 - 15 (s), respectively. In total, 180 reaction cycles were done. 

For the inhibited ALD experiments, methanol (CH$_3$OH, $\geq$99.8\%, Sigma-Aldrich) was used as an inhibitor and dosed prior to the DEZ pulse. An ice/$\text{CaCl}_2$ cooling bath was used to maintain the methanol precursor cylinder at approximately $-20\,^{\circ}\text{C}$ to reduce its vapor pressure. When methanol was used, the ALD sequence was MeOH pulse–purge–DEZ pulse–purge–H$_2$O pulse–purge, with the pulse and purge times being  0.2 - 10.8 - 0.2 - 15 - 0.2 - 15 (s), respectively. The pulse and purge times were chosen in a way such that the exposure of DEZ, and the overall cycle duration stayed constant for both the ALD runs, with and without the inhibitor.

\subsection{Conformality tests}
To analyze partial ALD conformality, rectangular
 lateral high-aspect-ratio (LHAR) test chips \, (PillarHall$^{\mathrm{TM}}$, generation LHAR5b) were used. Before ALD, in the reactor, an LHAR test chip was placed in addition to the Si wafer (the chip was on top of the wafer). Therefore, both the LHAR and the reference Si wafer were placed in the reactor in the same run. The test chip consisted of various test structures from which, for the analysis, we chose a mirrored test structure. The LHAR test structure had an open area (Zone I) from which reactants entered into the channel. The roof of the channel was made of a single-crystal silicon membrane, supported by SiO$_2$ pillars. After ALD, the channel roof was peeled off, making it possible to analyze the exposed film.  In the mirrored structure, the open area (Zone I) was at the center, with channels on either sides. The channel height was 500~nm, and the length was 1000~$\mu$m, corresponding to an aspect ratio of 2000:1. SEM images of the structure are shown in the supporting information (Figure S12). Calculated values of the Kn number for the experiments were 36.0, 36.4, and 37.2 for reaction temperatures of 150, 175, and 200~$^\circ\text{C}$, respectively, when calculated using Equation~\ref{eq:Kn_number}. Other parameters used for the Kn number calculation were: DEZ molar mass $M_A = 0.1235~\text{kg/mol}$, inert nitrogen gas molar mass $M_I = 0.028~\text{kg/mol}$, DEZ diameter $d_A = 6.90 \times 10^{-10}~\text{m}$ (calculated using Equation~8 in Ref.~\cite{ylilammi2018modeling}, with a DEZ density of $1200~\text{kg/m}^3 $\cite{haimi2026atomic}), nitrogen diameter $d_I = 3.74 \times 10^{-10}~\text{m}$, channel height $H = 500~\text{nm}$, channel width $W = 3300~\mu\text{m}$, partial pressure of reactant A $p_{A0}$ was the same as experimentally measured chamber pressures (0.92, 0.90, and 0.86 hPa) for experiments at 150, 175 and 200 $\degree$ C, and inert gas pressure $p_{I}$ was assumed to be 1 hPa.

\subsection{Thin film characterization}

\subsubsection{Ellipsometry}

On the ALD-coated Si wafers, the film thickness was determined using spectroscopic ellipsometry. Measurements were performed using a HORIBA UVISEL+ ellipsometer in the energy range of 1.5–4.5 eV with 0.05 eV increments at an incidence angle of 70$\degree$. The data was fitted using a model consisting of a ZnO layer described by a Tauc–Lorentz single-oscillator dispersion model on a SiO$_2$/c-Si substrate. The optical constants of the SiO$_2$ and c-Si layers were kept fixed using the HORIBA reference files SiO$_2$\_HJY.ref and c-Si\_HJY.ref, respectively.

\subsubsection{X-ray reflectometry}
On the ALD-coated Si wafers, film density and thickness were determined by X-ray reflectivity (XRR). Measurements were performed using a Bruker AXS D8 Advance diffractometer (Bruker AXS SE, Germany) equipped with a Cu K$\alpha$ radiation source ($\lambda$ = 1.5406) and an Eiger2 R 500K detector. The instrument was configured with a Göbel mirror to provide a parallel beam geometry. XRR scans were recorded in the angular range of approximately 0.1–6° with an appropriate step size and counting time. The reflectivity data were analyzed using the Parratt formalism implemented in DIFFRAC.LEPTOS X software, from which the film density, thickness, and roughness were extracted by fitting the experimental curves with a multilayer model.

\subsubsection{Scanning electron microscopy}
On the PillarHall$^{\mathrm{TM}}$ LHAR test chips, scanning electron microscopy (SEM) was performed using a Jeol JSM-IT800HL field emission SEM (FE-SEM) equipped with an Oxford energy dispersive X-ray spectrometer (EDS). Elemental distributions were characterized via X-ray line scans with a point spacing of approximately 1 $\mu$m. Line-scan repetitions were done three times to ensure data reliability. Additionally, elemental mapping was also conducted (Figure S12). An acceleration voltage of 5 keV was used for all measurements.  

Saturation profiles from the line-scans were converted to a Type-I normalized profile \cite{yim2020saturation} to compare penetration depth and slope. Normalization was done by dividing the X-ray signal by the maximum intensity, which was typically averaged over a distance of 20–60 $\mu$m within the channel. Only, for the standard ALD process (DEZ/water) at 150 $\degree$C, the maximum intensity was taken from the 10–20 $\mu$m range due to a sharp decrease in X-ray intensity at greater distances. Background subtraction was performed by referencing the signal at 500–700 $\mu$m, where the X-ray intensity reached its minimum.

\bibliography{Maindoc/references} 

@article{ylilammi2018modeling,
  title={Modeling growth kinetics of thin films made by atomic layer deposition in lateral high-aspect-ratio structures},
  author={Ylilammi, Markku and Ylivaara, Oili ME and Puurunen, Riikka L},
  journal={Journal of Applied Physics},
  volume={123},
  number={20},
  pages={205301},
  year={2018},
  publisher={AIP Publishing}
}

@article{richey2020understanding,
  title={Understanding chemical and physical mechanisms in atomic layer deposition},
  author={Richey, Nathaniel E and De Paula, Camila and Bent, Stacey F},
  journal={The Journal of Chemical Physics},
  volume={152},
  number={4},
  pages={040902},
  year={2020},
  publisher={AIP Publishing}
}

@article{sonsteby2020consistency,
  title={Consistency and reproducibility in atomic layer deposition},
  author={S{\o}nsteby, Henrik H and Yanguas-Gil, Angel and Elam, Jeffrey W},
  journal={Journal of Vacuum Science \& Technology A},
  volume={38},
  number={2},
  pages={020804},
  year={2020},
  publisher={AIP Publishing}
}

@article{kessels2025atomic,
  title={Atomic layer deposition},
  author={Kessels, Erwin and Devi, Anjana and Park, Jin-Seong and Ritala, Mikko and Yanguas-Gil, Angel and Wiemer, Claudia},
  journal={Nature Reviews Methods Primers},
  volume={5},
  number={1},
  pages={66},
  year={2025},
  publisher={Nature Publishing Group UK London}
}

@article{weber2023assessing,
  title={Assessing the environmental impact of atomic layer deposition (ALD) processes and pathways to lower it},
  author={Weber, Matthieu and Boysen, Nils and Graniel, Octavio and Sekkat, Abderrahime and Dussarrat, Christian and Wiff, Paulo and Devi, Anjana and Mu{\~n}oz-Rojas, David},
  journal={ACS materials Au},
  volume={3},
  number={4},
  pages={274},
  year={2023}
}

@article{piechulla2026atomic,
  title={Atomic layer deposition on particulate materials from 1988 through 2023: A quantitative review of technologies, materials, and applications},
  author={Piechulla, Peter M and Chen, Mingliang and Goulas, Aristeidis and Puurunen, Riikka L and van Ommen, J Ruud},
  journal={Chemistry of Materials},
  volume={38},
  number={1},
  pages={20--86},
  year={2026},
  publisher={ACS Publications}
}

@article{yim2022conformality,
  title={Conformality of atomic layer deposition in microchannels: impact of process parameters on the simulated thickness profile},
  author={Yim, Jihong and Verkama, Emma and Velasco, Jorge A and Arts, Karsten and Puurunen, Riikka L},
  journal={Physical Chemistry Chemical Physics},
  volume={24},
  number={15},
  pages={8645--8660},
  year={2022},
  publisher={Royal Society of Chemistry}
}

@article{gonsalves2024simulated,
  title={Simulated conformality of atomic layer deposition in lateral channels: the impact of the Knudsen number on the saturation profile characteristics},
  author={Gonsalves, Christine and Velasco, Jorge A and Yim, Jihong and J{\"a}rvilehto, J{\"a}nis and Vuorinen, Ville and Puurunen, Riikka L},
  journal={Physical Chemistry Chemical Physics},
  volume={26},
  number={45},
  pages={28431--28448},
  year={2024},
  publisher={Royal Society of Chemistry}
}

@article{kim2024enhancement,
  title={Enhancement of Conformality of Silicon Nitride Thin Films by ABC-Type Atomic Layer Deposition},
  author={Kim, Jiwon and Yeon, Changbong and Cho, Deok-Hyun and Jung, Jaesun and Shong, Bonggeun},
  journal={Advanced Electronic Materials},
  volume={10},
  number={3},
  pages={2300722},
  year={2024},
  publisher={Wiley Online Library}
}

@article{jeon2024synthesis,
  title={Synthesis of highly conformal titanium nitride films via tert-butyl chloride-assisted atomic layer deposition},
  author={Jeon, Jinho and Park, Heungsoo and Ko, Dae-Hong},
  journal={Applied Surface Science},
  volume={643},
  pages={158670},
  year={2024},
  publisher={Elsevier}
}

@article{yanguas2013modulation,
  title={Modulation of the growth per cycle in atomic layer deposition using reversible surface functionalization},
  author={Yanguas-Gil, Angel and Libera, Joseph A and Elam, Jeffrey W},
  journal={Chemistry of Materials},
  volume={25},
  number={24},
  pages={4849--4860},
  year={2013},
  publisher={ACS Publications}
}

@article{mameli2023selection,
  title={Selection criteria for small-molecule inhibitors in area-selective atomic layer deposition: fundamental surface chemistry considerations},
  author={Mameli, Alfredo and Teplyakov, Andrew V},
  journal={Accounts of Chemical Research},
  volume={56},
  number={15},
  pages={2084--2095},
  year={2023},
  publisher={ACS Publications}
}

@article{yim2023atomic,
  title={Atomic layer deposition of zinc oxide on mesoporous zirconia using zinc (II) acetylacetonate and air},
  author={Yim, Jihong and Haimi, Eero and Mantymaki, Miia and Karkas, Ville and Bes, Ren{\'e} and Gutierrez, Aitor Arandia and Meinander, Kristoffer and Br{\"u}ner, Philipp and Grehl, Thomas and Gell, Lars and others},
  journal={Chemistry of Materials},
  volume={35},
  number={19},
  pages={7915--7930},
  year={2023},
  publisher={ACS Publications}
}

@misc{velasco2026ald,
  author       = {Velasco, J. A. and Puurunen, R. L.},
  title        = {{ALD\_LHAR\_JV -- Diffusion-reaction model for ALD on a rectangular lateral high-aspect-ratio cavity (v1.8.1.1)}},
  year         = {2026},
  howpublished = {GitHub repository: \url{https://github.com/Aalto-Puurunen/ALD_LHAR_JV}},
  doi          = {10.5281/zenodo.22304150},
  url          = {https://doi.org/10.5281/zenodo.22304150}
}

@article{elam2003conformal,
  title={Conformal coating on ultrahigh-aspect-ratio nanopores of anodic alumina by atomic layer deposition},
  author={Elam, JW and Routkevitch, D and Mardilovich, PP and George, SM},
  journal={Chemistry of materials},
  volume={15},
  number={18},
  pages={3507--3517},
  year={2003},
  publisher={ACS Publications}
}

@article{lodha2024area,
  title={Area-selective atomic layer deposition of Ru using carbonyl-based precursor and oxygen co-reactant: Understanding defect formation mechanisms},
  author={Lodha, Jayant Kumar and Meersschaut, Johan and Pasquali, Mattia and Billington, Hans and Gendt, Stefan De and Armini, Silvia},
  journal={Nanomaterials},
  volume={14},
  number={14},
  pages={1212},
  year={2024},
  publisher={MDPI}
}

@article{kytokivi1997reaction,
  title={Reaction of acetylacetone vapour with [gamma]-alumina},
  author={Kyt{\"o}kivi, Arla and Rautiainen, Aimo and Root, Andrew},
  journal={Journal of the Chemical Society, Faraday Transactions},
  volume={93},
  number={22},
  pages={4079--4084},
  year={1997},
  publisher={The Royal Society of Chemistry}
}

@article{tynell2014atomic,
  title={Atomic layer deposition of ZnO: a review},
  author={Tynell, Tommi and Karppinen, Maarit},
  journal={Semiconductor Science and Technology},
  volume={29},
  number={4},
  pages={043001},
  year={2014},
  publisher={IOP Publishing}
}

@article{jensen2002x,
  title={X-ray reflectivity characterization of ZnO/Al$_2$O$_3$ multilayers prepared by atomic layer deposition},
  author={Jensen, JM and Oelkers, AB and Toivola, R and Johnson, David C and Elam, JW and George, SM},
  journal={Chemistry of Materials},
  volume={14},
  number={5},
  pages={2276--2282},
  year={2002},
  publisher={ACS Publications}
}

@article{sadruddin2026designing,
  title={Designing an agentic AI workflow for structured information extraction from scientific text: A case study on atomic layer deposition of ZnO and IGZO},
  author={Sadruddin, Sameer and Poupaki, Eleni and D’Souza, Jennifer and Auer, S{\"o}ren and Watkins, Alex and Karasulu, Bora and Mackus, Adriaan JM and Kessels, Erwin},
  journal={Journal of Vacuum Science \& Technology A},
  volume={44},
  number={3},
  pages={032413},
  year={2026},
  publisher={AIP Publishing}
}

@article{jeon2008structural,
  title={Structural and electrical properties of ZnO thin films deposited by atomic layer deposition at low temperatures},
  author={Jeon, Sunyeol and Bang, Seokhwan and Lee, Seungjun and Kwon, Semyung and Jeong, Wooho and Jeon, Hyeongtag and Chang, Ho Jung and Park, Hyung-Ho},
  journal={Journal of the Electrochemical Society},
  volume={155},
  number={10},
  pages={H738--H743},
  year={2008},
  publisher={The Electrochemical Society, Inc.}
}

@article{kim2011properties,
  title={The properties of plasma-enhanced atomic layer deposition (ALD) ZnO thin films and comparison with thermal ALD},
  author={Kim, Doyoung and Kang, Hyemin and Kim, Jae-Min and Kim, Hyungjun},
  journal={Applied Surface Science},
  volume={257},
  number={8},
  pages={3776--3779},
  year={2011},
  publisher={Elsevier}
}

@article{yarbrough2021next,
  title={Next generation nanopatterning using small molecule inhibitors for area-selective atomic layer deposition},
  author={Yarbrough, Josiah and Shearer, Alex B and Bent, Stacey F},
  journal={Journal of Vacuum Science \& Technology A},
  volume={39},
  number={2},
  pages={021002},
  year={2021},
  publisher={AIP Publishing}
}

@article{li2025competitive,
  title={Competitive Adsorption of Small Molecule Inhibitors and Trimethylaluminum Precursors on the Cu (111) Surface during Area-Selective Atomic Layer Deposition: A GCMC Study},
  author={Li, Chen and Li, Yichun and Weng, Jiayu and Chen, Jiafeng and Cao, Xiaoyong and Wei, Chunlei and Xu, Nan and He, Yi},
  journal={Langmuir},
  volume={41},
  number={4},
  pages={2572--2579},
  year={2025},
  publisher={ACS Publications}
}

@article{yu2024blocking,
  title={Blocking mechanisms in area-selective ALD by small molecule inhibitors of different sizes: Steric shielding versus chemical passivation},
  author={Yu, Pengmei and Merkx, Marc JM and Tezsevin, Ilker and Lemaire, Paul C and Hausmann, Dennis M and Sandoval, Tania E and Kessels, Wilhelmus MM and Mackus, Adriaan JM},
  journal={Applied Surface Science},
  volume={665},
  pages={160141},
  year={2024},
  publisher={Elsevier}
}

@article{shearer2024role,
  title={Role of molecular orientation: Comparison of nitrogenous aromatic small molecule inhibitors for area-selective atomic layer deposition},
  author={Shearer, Alexander and Pieck, Fabian and Yarbrough, Josiah and Werbrouck, Andreas and Tonner-Zech, Ralf and Bent, Stacey F},
  journal={Chemistry of Materials},
  volume={37},
  number={1},
  pages={139--152},
  year={2024},
  publisher={ACS Publications}
}

@article{tezsevin2023computational,
  title={Computational investigation of precursor blocking during area-selective atomic layer deposition using aniline as a small-molecule inhibitor},
  author={Tezsevin, I and Maas, JFW and Merkx, MJM and Lengers, R and Kessels, WMM and Sandoval, TE and Mackus, AJM},
  journal={Langmuir},
  volume={39},
  number={12},
  pages={4265},
  year={2023}
}

@article{reiter2024modeling,
  title={Modeling the impact of incomplete conformality during atomic layer processing},
  author={Reiter, Tobias and Aguinsky, Luiz Felipe and Rodrigues, Fr{\^a}ncio and Weinbub, Josef and H{\"o}ssinger, Andreas and Filipovic, Lado},
  journal={Solid-State Electronics},
  volume={211},
  pages={108816},
  year={2024},
  publisher={Elsevier}
}

@article{jung2022effect,
  title={Effect of hydrogen plasma treatment on atomic layer deposited silicon nitride film},
  author={Jung, Chanwon and Song, Seokhwi and Kim, Jisoo and Park, Suhyeon and Kim, Byunguk and Kim, Kyunghoo and Jeon, Hyeongtag},
  journal={ECS Journal of Solid State Science and Technology},
  volume={11},
  number={6},
  pages={063014},
  year={2022},
  publisher={IOP Publishing}
}

@article{nguyen2022gradient,
  title={Gradient area-selective deposition for seamless gap-filling in 3D nanostructures through surface chemical reactivity control},
  author={Nguyen, Chi Thang and Cho, Eun-Hyoung and Gu, Bonwook and Lee, Sunghee and Kim, Hae-Sung and Park, Jeongwoo and Yu, Neung-Kyung and Shin, Sangwoo and Shong, Bonggeun and Lee, Jeong Yub and others},
  journal={Nature Communications},
  volume={13},
  number={1},
  pages={7597},
  year={2022},
  publisher={Nature Publishing Group UK London}
}

@article{choolakkal2025using,
  title={Using a heavy inert diffusion additive for superconformal atomic layer deposition},
  author={Choolakkal, Arun Haridas and Mpofu, Pamburayi and Niiranen, Pentti and Birch, Jens and Pedersen, Henrik},
  journal={The Journal of Physical Chemistry Letters},
  volume={16},
  number={9},
  pages={2369--2372},
  year={2025},
  publisher={ACS Publications}
}

@article{cai2019revisit,
  title={A revisit to atomic layer deposition of zinc oxide using diethylzinc and water as precursors},
  author={Cai, Jiyu and Ma, Zhiyuan and Wejinya, Uche and Zou, Min and Liu, Yuzi and Zhou, Hua and Meng, Xiangbo},
  journal={Journal of Materials Science},
  volume={54},
  number={7},
  pages={5236--5248},
  year={2019},
  publisher={Springer}
}

@article{weckman2018atomic,
  title={Atomic layer deposition of zinc oxide: Study on the water pulse reactions from first-principles},
  author={Weckman, Timo and Laasonen, Kari},
  journal={The Journal of Physical Chemistry C},
  volume={122},
  number={14},
  pages={7685--7694},
  year={2018},
  publisher={ACS Publications}
}

@article{van2023excellent,
  title={Excellent conformality of atmospheric-pressure plasma-enhanced spatial atomic layer deposition with subsecond plasma exposure times},
  author={van de Poll, Mike L and Jain, Hardik and Hilfiker, James N and Utriainen, Mikko and Poodt, Paul and Kessels, Wilhelmus MM and Macco, Bart},
  journal={Applied Physics Letters},
  volume={123},
  number={18},
  pages={182902},
  year={2023},
  publisher={AIP Publishing}
}

@article{miikkulainen2013crystallinity,
  title={Crystallinity of inorganic films grown by atomic layer deposition: Overview and general trends},
  author={Miikkulainen, Ville and Leskel{\"a}, Markku and Ritala, Mikko and Puurunen, Riikka L},
  journal={Journal of Applied Physics},
  volume={113},
  number={2},
  pages={021301},
  year={2013},
  publisher={AIP Publishing}
}

@inproceedings{philip20233d,
  title={3D Thin Film Metrology without Cross-Sectional Sampling},
  author={Philip, Anish and Utriainen, Mikko and Werner, Thomas and Hyttinen, Pasi and Saarilahti, Jaakko and Kinnunen, Jussi and Gao, Feng},
  booktitle={2023 IEEE International Interconnect Technology Conference (IITC) and IEEE Materials for Advanced Metallization Conference (MAM)(IITC/MAM)},
  pages={1--3},
  year={2023},
  organization={IEEE}
}

@article{philip2024conformal,
  title={Conformal Zn-Benzene Dithiol Thin Films for Temperature-Sensitive Electronics Grown via Industry-Feasible Atomic/Molecular Layer Deposition Technique},
  author={Philip, Anish and Jussila, Topias and Obenl{\"u}neschlo{\ss}, Jorit and Zanders, David and Preischel, Florian and Kinnunen, Jussi and Devi, Anjana and Karppinen, Maarit},
  journal={Small},
  volume={20},
  number={40},
  pages={2402608},
  year={2024},
  publisher={Wiley Online Library}
}

@article{heikkinen2024atomic,
  title={An atomic layer deposition diffusion--reaction model for porous media with different particle geometries},
  author={Heikkinen, Niko and Lehtonen, Juha and Puurunen, Riikka L},
  journal={Physical Chemistry Chemical Physics},
  volume={26},
  number={9},
  pages={7580--7591},
  year={2024},
  publisher={Royal Society of Chemistry}
}

@article{janocha2011ald,
  title={ALD of ZnO using diethylzinc as metal-precursor and oxygen as oxidizing agent},
  author={Janocha, E and Pettenkofer, C},
  journal={Applied Surface Science},
  volume={257},
  number={23},
  pages={10031--10035},
  year={2011},
  publisher={Elsevier}
}

@article{yanguas2012self,
  title={Self-Limited Reaction-Diffusion in Nanostructured Substrates: Surface Coverage Dynamics and Analytic Approximations to ALD Saturation Times},
  author={Yanguas-Gil, Angel and Elam, Jeffrey W},
  journal={Chemical Vapor Deposition},
  volume={18},
  number={1-3},
  pages={46--52},
  year={2012},
  publisher={Wiley Online Library}
}

@article{puurunen2003growth,
  title={Growth per cycle in atomic layer deposition: a theoretical model},
  author={Puurunen, Riikka L},
  journal={Chemical Vapor Deposition},
  volume={9},
  number={5},
  pages={249--257},
  year={2003},
  publisher={Wiley Online Library}
}

@article{puurunen2003growththeot,
  title={Growth per cycle in atomic layer deposition: real application examplesof a theoretical model},
  author={Puurunen, Riikka L},
  journal={Chemical Vapor Deposition},
  volume={9},
  number={6},
  pages={327--332},
  year={2003},
  publisher={Wiley Online Library}
}

@article{haimi2026atomic,
  title={Atomic layer deposition of zinc oxide films on lateral high-aspect-ratio test structures using diethylzinc and water as precursors},
  author={Haimi, Eero and Philip, Anish and Velasco, Jorge A and Gao, Feng and Karppinen, Maarit and Puurunen, Riikka L},
  journal={Journal of Vacuum Science \& Technology A},
  volume={44},
  number={2},
  pages={022410},
  year={2026},
  publisher={AIP Publishing}
}

@article{gordon2003kinetic,
  title={A kinetic model for step coverage by atomic layer deposition in narrow holes or trenches},
  author={Gordon, Roy G and Hausmann, Dennis and Kim, Esther and Shepard, Joseph},
  journal={Chemical Vapor Deposition},
  volume={9},
  number={2},
  pages={73--78},
  year={2003},
  publisher={Wiley Online Library}
}

@article{ismaeel2025area,
  title={Area selective atomic layer deposition of ruthenium with phenol as a small molecule inhibitor},
  author={Ismaeel, Sundas and Chundak, Mykhailo and Ritala, Mikko},
  journal={Journal of Vacuum Science \& Technology A},
  volume={43},
  number={6},
  year={2025},
  pages={062407},
  publisher={AIP Publishing}
}

@article{lee2025molecular,
  title={Molecular Design in Area-Selective Atomic Layer Deposition: Understanding Inhibitors and Precursors},
  author={Lee, Yujin and Rothman, Amnon and Shearer, Alexander B and Bent, Stacey F},
  journal={Chemistry of Materials},
  volume={37},
  number={5},
  pages={1741--1758},
  year={2025},
  publisher={ACS Publications}
}

@article{ham2022investigation,
  title={Investigation of abnormally high growth-per-cycle in atomic layer deposition of Al2O3 using trimethylaluminum and water},
  author={Ham, So-Yeon and Jin, Zhenyu and Shin, Seokhee and Kim, Minseo and Seo, Mingyu and Min, Yo-Sep},
  journal={Applied Surface Science},
  volume={571},
  pages={151282},
  year={2022},
  publisher={Elsevier}
}

@incollection{van_ommen_atomic_2021,
	title = {Atomic {Layer} {Deposition}},
	
	url = {https://onlinelibrary.wiley.com/doi/abs/10.1002/0471238961.koe00059},
	
	booktitle = {Kirk-{Othmer} {Encyclopedia} of {Chemical} {Technology}},
	publisher = {John Wiley \& Sons, Ltd},
	author = {van Ommen, J. Ruud and Goulas, Aristeidis and Puurunen, Riikka L.},
	year = {2021},
	doi = {10.1002/0471238961.koe00059},
	
	
	pages = {1--42},
	
}

@article{cremers2019conformality,
  title={Conformality in atomic layer deposition: Current status overview of analysis and modelling},
  author={Cremers, V{\'e}ronique and Puurunen, Riikka L and Dendooven, Jolien},
  journal={Applied Physics Reviews},
  volume={6},
  number={2},
  pages={021302},
  year={2019},
  publisher={AIP Publishing}
}

@article{jo2026advances,
  title={Advances in area-selective atomic layer deposition: surface chemistry, challenges, and future perspectives},
  author={Jo, Hyosik and Kim, Yunseok and Choi, Seulwon and Yoo, Ilhan and Han, Minji and Ryu, Jung-El and Park, Hwanyeol},
  journal={Coordination Chemistry Reviews},
  volume={558},
  pages={217768},
  year={2026},
  publisher={Elsevier}
}

@article{oh2021role,
  title={Role of precursor choice on area-selective atomic layer deposition},
  author={Oh, Il-Kwon and Sandoval, Tania E and Liu, Tzu-Ling and Richey, Nathaniel E and Bent, Stacey F},
  journal={Chemistry of Materials},
  volume={33},
  number={11},
  pages={3926--3935},
  year={2021},
  publisher={ACS Publications}
}

@article{merkx2022relation,
  title={Relation between reactive surface sites and precursor choice for area-selective atomic layer deposition using small molecule inhibitors},
  author={Merkx, Marc JM and Angelidis, Athanasios and Mameli, Alfredo and Li, Jun and Lemaire, Paul C and Sharma, Kashish and Hausmann, Dennis M and Kessels, Wilhelmus MM and Sandoval, Tania E and Mackus, Adriaan JM},
  journal={The Journal of Physical Chemistry C},
  volume={126},
  number={10},
  pages={4845--4853},
  year={2022},
  publisher={ACS Publications}
}

@article{puurunen_surface_2005,
	title = {Surface chemistry of atomic layer deposition: {A} case study for the trimethylaluminum/water process},
author = {Puurunen, Riikka L.},
journal = {Journal of Applied Physics},
year = {2005},
pages ={121301},
	volume = {97},
number = {12},

	
	url = {https://pubs.aip.org/aip/jap/article/97/12/121301/893976/Surface-chemistry-of-atomic-layer-deposition-A},
	doi = {10.1063/1.1940727},
	
}

@article{tanaka2025growth,
  title={Growth of Bismuth Oxide by Atomic Layer Deposition: An Attempt to Achieve a Stoichiometric Composition and a High Growth Rate},
  author={Tanaka, Yu and Sakama, Hiroshi},
  journal={Crystal Growth \& Design},
  volume={25},
  number={4},
  pages={970--977},
  year={2025},
  publisher={ACS Publications}
}

@article{zhang2019high,
  title={High growth per cycle thermal atomic layer deposition of Ni films using an electron-rich precursor},
  author={Zhang, Yuxiang and Du, Liyong and Liu, Xinfang and Ding, Yuqiang},
  journal={Nanoscale},
  volume={11},
  number={8},
  pages={3484--3488},
  year={2019},
  publisher={Royal Society of Chemistry}
}

@inproceedings{puurunen2016influence,
  title={Influence of ALD temperature on thin film conformality: Investigation with microscopic lateral high-aspect-ratio structures},
  author={Puurunen, Riikka L and Gao, Feng},
  booktitle={Influence of ALD temperature on thin film conformality: Investigation with microscopic lateral high-aspect-ratio structures, 14th International Baltic Conference on Atomic Layer Deposition (BALD)},
  pages={20--24},
  year={2016},
  publisher={IEEE Institute of Electrical and Electronic Engineers}
}

@article{yim2020saturation,
  title={Saturation profile based conformality analysis for atomic layer deposition: aluminum oxide in lateral high-aspect-ratio channels},
  author={Yim, Jihong and Ylivaara, Oili ME and Ylilammi, Markku and Korpelainen, Virpi and Haimi, Eero and Verkama, Emma and Utriainen, Mikko and Puurunen, Riikka L},
  journal={Physical Chemistry Chemical Physics},
  volume={22},
  number={40},
  pages={23107--23120},
  year={2020},
  publisher={Royal Society of Chemistry}
}

@article{mattinen2016nucleation,
  title={Nucleation and conformality of iridium and iridium oxide thin films grown by atomic layer deposition},
  author={Mattinen, Miika and Hamalainen, Jani and Gao, Feng and Jalkanen, Pasi and Mizohata, Kenichiro and Raisanen, Jyrki and Puurunen, Riikka L and Ritala, Mikko and Leskela, Markku},
  journal={Langmuir},
  volume={32},
  number={41},
  pages={10559--10569},
  year={2016},
  publisher={ACS Publications}
}

@article{arts_sticking_2019,
    title = {Sticking probabilities of {H2O} and {Al}({CH3})3 during atomic layer deposition of {Al2O3} extracted from their impact on film conformality},
    volume = {37},
    issn = {0734-2101, 1520-8559},
    url = {https://pubs.aip.org/jva/article/37/3/030908/910143/Sticking-probabilities-of-H2O-and-Al-CH3-3-during},
    doi = {10.1116/1.5093620},
    language = {en},
    number = {3},
    urldate = {2026-01-28},
    journal = {Journal of Vacuum Science \& Technology A: Vacuum, Surfaces, and Films},
    author = {Arts, Karsten and Vandalon, Vincent and Puurunen, Riikka L. and Utriainen, Mikko and Gao, Feng and Kessels, Wilhelmus M. M. (Erwin) and Knoops, Harm C. M.},
    month = may,
    year = {2019},
    pages = {030908},
}

@article{jarvilehto_simulation_2023,
    title = {Simulation of conformality of {ALD} growth inside lateral channels: comparison between a diffusion–reaction model and a ballistic transport–reaction model},
    volume = {25},
    issn = {1463-9084},
    shorttitle = {Simulation of conformality of {ALD} growth inside lateral channels},
    url = {https://pubs.rsc.org/en/content/articlelanding/2023/cp/d3cp01829f},
    doi = {10.1039/D3CP01829F},
    language = {en},
    number = {34},
    urldate = {2025-05-19},
    journal = {Physical Chemistry Chemical Physics},
    publisher = {The Royal Society of Chemistry},
    author = {Järvilehto, Jänis and Velasco, Jorge A. and Yim, Jihong and Gonsalves, Christine and Puurunen, Riikka L.},
    month = aug,
    year = {2023},
    pages = {22952--22964},
}
\bibliographystyle{rsc} 

\end{document}


\vspace{4cm}
\noindent \large{\textbf{Electronic supplementary information}}\\

\noindent\LARGE{\textbf{Enhancing Conformality in Atomic Layer Deposition through Low Growth Per Cycle}}\\

\noindent\large{
Christine Gonsalves\textit{*$^{a}$}, Jorge A. Velasco\textit{$^{a}$}, Ahmed Othman\textit{$^{b}$}, Ville Miikkulainen\textit{$^{b}$}, and  Riikka L. Puurunen\textit{*$^{a}$}

\let\thefootnote\relax\footnote{\textit{$^{a}$~Department of Chemical and Metallurgical Engineering, Aalto University,
P.O. Box 16100, FI-00076 AALTO, Finland.}}
\let\thefootnote\relax\footnote{\textit{$^{b}$~Department of Chemistry and Materials Science, Aalto University,
P.O. Box 16100, FI-00076 AALTO, Finland.}}

\footnotetext{* Corresponding authors. E-mail: christine.gonsalves@aalto.fi, riikka.puurunen@aalto.fi}

\vspace{1em}

\tableofcontents

\clearpage

\section{Literature summary}
Analysis of existing literature of film penetration in HAR structures, results in the observation that film penetration depth generally \textit{increases} as the ALD growth per cycle \textit{decreases} (Table \ref{tab:ald_summary}). Examples from experimental studies of $\text{Al}_2\text{O}_3$ and $\text{TiO}_2$ films in lateral high-aspect-ratio structures (LHAR) structures showed that when the GPC decreased as a result of changing the reaction temperature, the film's penetration depth increased\cite{puurunen2016influence}. In another case for $\text{Al}_2\text{O}_3$, when the GPC decreased as a result of changing the purge times, the penetration depth increased\cite{yim2020saturation}. For $\text{IrO}_2$ films, when the GPC decreased as a result of changing the second reactant, the penetration depth increased\cite{mattinen2016nucleation}.  Also, in modeling studies, it was observed that the film penetration depth increased with a decrease in the GPC~\cite{gonsalves2024simulated,yim2022conformality}. Additionally, analysis of works that evaluate ALD conformality in terms of \textit{step coverage} (ratio of film thickness at the bottom to the top)\cite{cremers2019conformality}, also show that when the GPC decreased, step coverage increased (Table \ref{tab:ald_stepcoverage})~\cite{kim2024enhancement,jeon2024synthesis,nguyen2022gradient,jung2022effect}. In all the studies, the process parameter changes that resulted in increased conformality were accompanied with a decrease in the GPC~\cite{puurunen2016influence,mattinen2016nucleation,yim2020saturation,yim2022conformality,gonsalves2024simulated,kim2024enhancement,jeon2024synthesis,nguyen2022gradient,jung2022effect}. The details of the correlation between GPC and film penetration depth have remained poorly described and understood and will be explored in this study.

By simulations, GPC can be easily varied but varying GPC intentionally experimentally is less straightforward. To intentionally and systematically decrease the GPC in experiments, one approach could be to use inhibitor molecules. Inhibitor molecules are commonly used in area-selective ALD \cite{lee2025molecular,shearer2024role,mameli2023selection,ismaeel2025area} to block available surface sites to create a non-growth surface \cite{yu2024blocking,merkx2022relation,oh2021role,jo2026advances}. A schematic of inhibitor related site blocking is shown in Figure 1c of the main article. Inhibitors have also been shown to successfully decrease the GPC of many ALD processes \cite{yanguas2013modulation,li2025competitive,merkx2022relation}. Examples include using alcohols, organic acids (ethanol, methanol, carboxylic acid, etc.)\cite{yu2024blocking,li2025competitive,yarbrough2021next,merkx2022relation} and $\beta$-diketones like acetylacetone (Hacac) \cite{kytokivi1997reaction} to block the reaction of common ALD alkyl or halide precursors ($\text{TiCl}_4$, TMA, DEZ, etc.), resulting in GPC values that are reduced by half or suppressed entirely.\cite{yanguas2013modulation,li2025competitive,merkx2022relation,lodha2024area,tezsevin2023computational,shearer2024role}
Inhibitors have recently also been used to increase ALD conformality in terms of the \textit{step-coverage}\cite{jeon2024synthesis,kim2011properties,nguyen2022gradient} meaning the ratio of film thickness at the bottom to the top of a structure\cite{cremers2019conformality} (Table \ref{tab:ald_stepcoverage}).  
 
\begin{sidewaystable}[p]
\centering
\scriptsize
\setlength{\tabcolsep}{3pt}
\renewcommand{\arraystretch}{1.3}

\caption{Summary of literature showing decrease in GPC resulted in increase in ALD film penetration depth in HAR structures.\textsuperscript{a,b,c}}
\label{tab:ald_summary}

\begin{tabular}{|>{\raggedright\arraybackslash}p{1.1cm}
                |>{\raggedright\arraybackslash}p{1.4cm}
                |>{\raggedright\arraybackslash}p{1.1cm}
                |>{\raggedright\arraybackslash}p{2.0cm}
                |>{\raggedright\arraybackslash}p{2.1cm}
                |>{\raggedright\arraybackslash}p{1.1cm}
                |>{\raggedright\arraybackslash}p{1.7cm}
                |>{\raggedright\arraybackslash}p{1.6cm}
                |>{\raggedright\arraybackslash}p{1.6cm}
                |>{\raggedright\arraybackslash}p{1.4cm}
                |>{\raggedright\arraybackslash}p{1.7cm}
                |>{\raggedright\arraybackslash}p{0.7cm}|}
\hline
\textbf{Material} & 
\textbf{Study Type} & 
\textbf{Temp. (\textdegree C)} & 
\textbf{Reactants} & 
\textbf{Sequence (s)}\newline pulse/purge*/ \newline pulse/purge* & 
\textbf{ALD cycles} & 
\textbf{Factor influencing GPC} & 
\textbf{GPC range (nm)} & 
\textbf{Penetration depth range ($\mu$m)} & 
\textbf{\% Decrease in GPC\textsuperscript{d}} & 
\textbf{\% Increase in penetration depth\textsuperscript{e}} & 
\textbf{Ref.} \\
\hline

TiO$_2$ & 
Experimental & 
110 to 300 & 
\ce{TiCl4} / \ce{H2O} & 
0.1/4*/0.1/4* & 
200 to 967 & 
Reaction temperature & 
0.05...0.04 & 
65...100 & 
upto 20\% & 
upto 54\% & 
\cite{puurunen2016influence} \\
\hline

Al$_2$O$_3$ & 
Experimental & 
110 to 300 & 
\ce{Al(CH3)3} / \ce{H2O} & 
0.1/4*/0.1/4* & 
514 to 626 & 
Reaction temperature & 
0.090...0.078 & 
105...130 & 
upto 19\% & 
upto 18\% & 
\cite{puurunen2016influence} \\
\hline

Ir, IrO$_2$ & 
Experimental & 
185 to 250 & 
\ce{Ir(acac)3} / \ce{O3}, \ce{O2}, \ce{H2} & 
2/5*/2/5* & 
1500 to 2500 & 
Second reactant & 
0.036...0.022 & 
17...44 & 
upto 39\% & 
upto 159\% & 
\cite{mattinen2016nucleation} \\
\hline

Al$_2$O$_3$ & 
Experimental & 
300 & 
\ce{Al(CH3)3} / \ce{H2O} & 
0.1/1,4,10*/ 0.1/1,4,10* & 
500 & 
Purge time & 
0.101...0.0897 & 
96...109 & 
upto 11\% & 
upto 10\% & 
\cite{yim2020saturation} \\
\hline

Al$_2$O$_3$ & 
Simulated & 
300 & 
\ce{Al(CH3)3} / \ce{H2O} & 
0.1/*//*$^f$ & 
1 & 
Adsorption capacity & 
0.190...0.012 & 
12105...49117 & 
upto 94\% & 
upto 306\% & 
\cite{gonsalves2024simulated} \\
\hline

Al$_2$O$_3$ & 
Simulated & 
300 & 
\ce{Al(CH3)3} / \ce{H2O} & 
0.1/*//*$^f$ & 
1 & 
Adsorption capacity & 
0.190...0.012 & 
6.5...26.5 & 
upto 94\% & 
upto 308\% & 
\cite{yim2022conformality} \\
\hline

\end{tabular}

\vspace{0.4em}
\begin{flushleft}
\scriptsize
\begin{itemize}[leftmargin=1.5em, itemsep=1pt, parsep=0pt, topsep=2pt]
    \item[\textsuperscript{a}] Authors' interpretation of the literature data.
    \item[\textsuperscript{b}] Detailed sample-wise analysis of the decrease in GPC and increase in film penetration is in Table \ref{tab:ald_penetration_depth}.
    \item[\textsuperscript{c}] All reported data correspond to LHAR test structures with aspect ratios up to 10,000:1.\cite{puurunen2016influence,mattinen2016nucleation,yim2020saturation,yim2022conformality,gonsalves2024simulated}
    \item[\textsuperscript{d}] Percentage decrease calculated as: $\frac{\text{Value}_1 - \text{Value}_i}{\text{Value}_1} \times 100\%$, where $\text{Value}_1$ is the value for Sample 1, and $\text{Value}_i$ is the value for Sample 2 or Sample 3.
    \item[\textsuperscript{e}] Percentage increase calculated as: $\frac{\text{Value}_i - \text{Value}_1}{\text{Value}_1} \times 100\%$, where $\text{Value}_1$ is the value for Sample 1, and $\text{Value}_i$ is the value for Sample 2 or Sample 3.
    \item[\textsuperscript{f}] Only the first ALD step: reactant pulse was simulated.\cite{gonsalves2024simulated,yim2022conformality}
    \item[*] Asterisks are used as visual placeholders for the purge steps.
\end{itemize}
\end{flushleft}

\end{sidewaystable}

\begin{sidewaystable}[p]
\centering
\scriptsize
\setlength{\tabcolsep}{2pt} 
\renewcommand{\arraystretch}{1.3}

\caption{Summary of literature, with sample-wise analysis where changes in GPC resulted in changes in ALD film penetration depth in HAR structures.$^a$}
\label{tab:ald_penetration_depth}

\begin{tabularx}{\linewidth}{|>{\raggedright\arraybackslash}p{1.1cm}|>{\raggedright\arraybackslash}p{1.1cm}|>{\raggedright\arraybackslash}X|>{\raggedright\arraybackslash}p{0.8cm}|>{\raggedright\arraybackslash}p{1.6cm}|>{\raggedright\arraybackslash}p{1.1cm}|>{\raggedright\arraybackslash}p{1.1cm}|>{\raggedright\arraybackslash}p{1.1cm}|c|c|c|c|c|c|c|c|c|c|p{0.7cm}|}
\hline
\textbf{Material, Temperature} & 
\textbf{Test structure} & 
\textbf{Reactants and Sequence} \newline pulse/purge*/\newline pulse/purge* (s) & 
\textbf{ALD cycles} & 
\textbf{Factor influencing GPC} & 
\textbf{Sample 1 (S1)} & 
\textbf{Sample 2 (S2)} & 
\textbf{Sample 3 (S3)} & 
\multicolumn{3}{c|}{\textbf{GPC (nm)}} & 
\multicolumn{3}{c|}{\textbf{Penetration depth ($\mu$m)}} & 
\multicolumn{2}{c|}{\makecell{\textbf{\% Decrease in}\\\textbf{GPC ($\downarrow$)}}$^b$} & 
\multicolumn{2}{c|}{\makecell{\textbf{\% Increase in}\\\textbf{penetration depth ($\uparrow$)}}$^c$} & 
\textbf{Ref.} \\
\cline{9-18}
&&&&&&&& \textbf{S1$^d$} & \textbf{S2} & \textbf{S3} & \textbf{S1} & \textbf{S2} & \textbf{S3} & \textbf{S1-to-S2} & \textbf{S1-to-S3} & \textbf{S1-to-S2} & \textbf{S1-to-S3} & \\
\hline

TiO$_2$, 110\,°C & 
LHAR, AR 10000:1 & 
\ce{TiCl4} / \ce{H2O} \newline 0.1/4*/0.1/4* & 
967 (S1)\newline 200 (S2)\newline 300 (S3) & 
Reaction temperature  & 
Process at 110\,°C & 
Process at 200\,°C & 
Process at 300\,°C & 
0.05 & 0.04 & 0.05 & 
65 & 100 & 65 & 
20.0\% & 0.0\% & 53.8\% & 0.0\% & 
\cite{puurunen2016influence} \\
\hline

Al$_2$O$_3$, 110\,°C & 
LHAR, AR 10000:1 & 
\ce{Al(CH3)3} / \ce{H2O} \newline 0.1/4*/0.1/4* & 
514 (S1)\newline 626 (S2)\newline 552 (S3) & 
Reaction temperature & 
Process at 200\,°C & 
Process at 110\,°C & 
Process at 300\,°C & 
0.096 & 0.078 & 0.090 & 
110 & 130 & 105 & 
18.8\% & 6.3\% & 18.2\% & -4.5\% & 
\cite{puurunen2016influence} \\
\hline

Ir, IrO$_2$, 185-250\,°C & 
LHAR, AR 10000:1 & 
\ce{Ir(acac)3} / \ce{O3}, \ce{H2O} \newline 2/5*/2/5* & 
1500 (S1)\newline 2500 (S2)\newline 1900 (S3) & 
Identity of reactant B & 
\ce{O2} process & 
\ce{O3} + \ce{H2} process & 
\ce{O2} + \ce{H2} process & 
0.036 & 0.022 & 0.026 & 
17 & 44 & 22 & 
38.9\% & 27.8\% & 158.8\% & 29.4\% & 
\cite{mattinen2016nucleation} \\
\hline

Al$_2$O$_3$, 300\,°C & 
LHAR, AR 10000:1 & 
\ce{Al(CH3)3} / \ce{H2O} \newline 0.1/1,4,10*/0.1/1,4,10* & 
500 & 
Purge time & 
Purge  = 1 s & 
Purge  = 10 s & 
Purge  = 4 s & 
0.101 & 0.0897 & 0.0951 & 
99 & 109 & 96 & 
11.2\% & 5.8\% & 10.1\% & -3.0\% & 
\cite{yim2020saturation} \\
\hline

Al$_2$O$_3$, 300\,°C & 
LHAR, AR 1000:1$^e$ & 
\ce{Al(CH3)3} / \ce{H2O} \newline 0.1/*//*$^f$ & 
1 & 
Adsorption \newline capacity & 
$8\,\text{nm}^{-2}$ & 
$0.5\,\text{nm}^{-2}$ & 
$4\,\text{nm}^{-2}$ & 
0.1898$^g$ & 0.0119$^g$ & 0.0949$^g$ & 
12105 & 49117 & 17215 & 
93.8\% & 50.0\% & 305.7\% & 42.2\% & 
\cite{gonsalves2024simulated} \\
\hline

Al$_2$O$_3$, 300\,°C & 
LHAR, AR 1000:1$^e$ & 
\ce{Al(CH3)3} / \ce{H2O} \newline 0.1/*//*$^f$  & 
1 & 
Adsorption \newline capacity & 
$8\,\text{nm}^{-2}$ & 
$0.5\,\text{nm}^{-2}$ & 
$4\,\text{nm}^{-2}$ & 
0.1898$^g$ & 0.0119$^g$ & 0.0949$^g$ & 
6.5 & 26.5 & 9.3 & 
93.8\% & 50.0\%& 307.7\% & 42.3\% & 
\cite{yim2022conformality} \\
\hline

\end{tabularx}

\vspace{0.4em}
\begin{flushleft}
\scriptsize \begin{itemize}[leftmargin=1.5em, itemsep=1pt, parsep=0pt, topsep=2pt]
    \item[$^a$] Authors' interpretation of the literature data.
    \item[$^b$] Percentage decrease calculated as: $\frac{\text{Value}_1 - \text{Value}_i}{\text{Value}_1} \times 100\%$, where $\text{Value}_1$ is the value for sample 1, and $\text{Value}_i$ is the value for sample 2 or sample 3.
    
    \item[$^c$] Percentage increase calculated as: $\frac{\text{Value}_i - \text{Value}_1}{\text{Value}_1} \times 100\%$, where $\text{Value}_1$ is the value for sample 1, and $\text{Value}_i$ is the value for sample 2 or sample 3. 
   \item[$^d$] Samples are ordered such that sample 1 corresponds to the sample with the highest GPC among the evaluated samples and is used as reference for comparisons.
    \item[$^e$] Data is from simulated saturation profiles\cite{gonsalves2024simulated,yim2022conformality}.
    \item[$^f$] Only the first ALD step: reactant pulse was simulated.\cite{gonsalves2024simulated,yim2022conformality}
    \item[$^g$] Calculated from the value of adsorption capacity (8/0.5/4 \#/nm$^2$) using Equation 1 of the main manuscript with values of $M$ = 0.050 kg/mol, $\rho$ = 3500 kg/m$^3$\cite{gonsalves2024simulated,yim2022conformality}.
    \item[*]Asterisks are used as visual placeholders for the purge steps.
\end{itemize}
\end{flushleft}

\end{sidewaystable}

\begin{sidewaystable}[p]
\centering
\scriptsize
\setlength{\tabcolsep}{2.5pt} 
\renewcommand{\arraystretch}{1.3}

\caption{Summary of literature data with sample-wise analysis where changes in GPC affected the ALD step coverage.$^a$}
\label{tab:ald_stepcoverage}

\begin{tabularx}{\linewidth}{|>{\raggedright\arraybackslash}p{1.1cm}|>{\raggedright\arraybackslash}p{1.2cm}|>{\raggedright\arraybackslash}p{3.2cm}|>{\raggedright\arraybackslash}p{0.9cm}|>{\raggedright\arraybackslash}p{1.3cm}|>{\raggedright\arraybackslash}p{1.3cm}|>{\raggedright\arraybackslash}p{1.3cm}|>{\raggedright\arraybackslash}p{1.3cm}|c|c|c|c|c|c|X|X|X|X|p{0.8cm}|}
\hline
\textbf{Material, Temperature} & 
\textbf{Test Structure} & 
\textbf{Reactants and Sequence} \newline reactant/purge*/\newline reactant/purge*/\newline reactant/purge* (s) & 
\textbf{ALD Cycles} & 
\textbf{Factor \newline influencing \newline GPC} & 
\textbf{Sample 1 (S1)} & 
\textbf{Sample 2 (S2)} & 
\textbf{Sample 3 (S3)} & 
\multicolumn{3}{c|}{\textbf{GPC (nm)}} & 
\multicolumn{3}{c|}{\textbf{Step Coverage $^b$ (\%)}} & 
\multicolumn{2}{c|}{\makecell{\textbf{\% Decrease in}\\\textbf{GPC ($\downarrow$)}}$^c$} & 
\multicolumn{2}{c|}{\makecell{\textbf{\% Increase in}\\\textbf{step coverage ($\uparrow$)}}$^d$} & 
\textbf{Ref.} \\
\cline{9-18}
&&&&&&&& \textbf{S1$^e$} & \textbf{S2} & \textbf{S3} & \textbf{S1} & \textbf{S2} & \textbf{S3} & \textbf{S1-to-S2} & \textbf{S1-to-S3} & \textbf{S1-to-S2} & \textbf{S1-to-S3} & \\
\hline

SiN$_x$, 400\,°C & 
3D struct., AR 2.5:1 & 
DTDN--\ce{H2}$^f$ / \ce{N2} plasma / \ce{H2} plasma \newline 30/30*/30/30*/10/10* & 
200 & 
Hydrogen treatment & 
no \ce{H2} plasma (baseline \ce{N2}) & 
\ce{H2} plasma (1:1) & 
\ce{H2} plasma (10:1) & 
0.036 & 0.032 & 0.035 & 
86\% & 98\% & 93.8\% & 
11.1\% & 2.8\% & 14.0\% & 9.1\% & 
\cite{jung2022effect} \\
\hline

SiN$_x$, 600\,°C & 
Vertical trench, AR 22:1 & 
TBC$^f$ / \ce{Si2Cl6} / \ce{NH3} \newline 0,5,10/*$^f$/30/30*/30/30* & 
91$^h$ (S1) \newline 200$^h$ (S2)\newline 167$^h$ (S3) & 
Inhibition & 
0 s TBC & 
10 s TBC & 
5 s TBC & 
0.11 & 0.05 & 0.06 & 
88\% & 122\% & 105\% & 
54.5\% & 45.5\% & 38.6\% & 19.3\% & 
\cite{kim2024enhancement} \\
\hline

TiN, 460\,°C & 
Patterned wafer, AR 20:1 & 
TBC$^f$ / \ce{TiCl4} / \ce{NH3} \newline 0.05/0.1*/0.05/0.2*/0.5/0.3* & 
444$^i$ (S1)\newline 860$^i$ (S2)\newline 1096$^i$ (S3) & 
Inhibition & 
0 sccm TBC & 
100 sccm TBC & 
25 sccm TBC & 
0.036 & 0.0146 & 0.0186 & 
82.8\% & 99.4\%& 98.9\% & 
59.4\% & 48.3\% & 20.0\% & 19.4\% & 
\cite{jeon2024synthesis} \\
\hline

TiO$_2$, 180\,°C & 
3D hole (d: 1600\,nm, open: 100\,nm, btm: 50\,nm) & 
TMPMCT$^f$ / TDMAT$^f$ / \ce{H2O} \newline 0,40,60/*//*/*/*$^g$ & 
730 & 
Inhibition & 
0 s TMPMCT & 
60 s TMPMCT & 
40 s TMPMCT & 
0.055 & 0.0027$^j$ & 0.014$^k$ & 
75\% & 1500\% & 300\% & 
95.1\% & 74.5\% & 1900.0\% & 300.0\% & 
\cite{nguyen2022gradient} \\
\hline

\end{tabularx}

\vspace{0.4em}
\begin{flushleft}
\scriptsize \begin{itemize}[leftmargin=1.5em, itemsep=1pt, parsep=0pt, topsep=2pt] 
$^a$ Authors' interpretation of the literature data.

$^b$ Bottom step coverage was calculated as the ratio of film thickness at the top to the film thickness at the bottom of the structure.

$^c$Percentage increase calculated as: $\frac{\text{Value}_i - \text{Value}_1}{\text{Value}_1} \times 100$\%, where $\text{Value}_1$ is the value for sample 1, and $\text{Value}_i$ is the value for sample 2 or sample 3.

$^d$Percentage decrease calculated as: $\frac{\text{Value}_1 - \text{Value}_i}{\text{Value}_1} \times 100$\%, where $\text{Value}_1$ is the value for sample 1, and $\text{Value}_i$ is the value for sample 2 or sample 3.

$^e$ Samples are ordered such that sample 1 corresponds to the sample with the highest GPC among the evaluated samples and is used as reference for comparisons.

$^f$ Abbreviations for: (DTDN-2H2) = bis(dimethylaminomethylsilyl)trimethylsilylamine, TBC = tert-butylchloride, TMPMCT=trimethoxy(pentamethylcyclopentadienyl)-titanium(IV), TDMAT=tetrakis(dimethylamido)titanium.

$^g$ Time not explicitly mentioned\cite{nguyen2022gradient}.

$^h$ Number of cycles was calculated based on the reported GPC and target film thickness (10 nm)\cite{kim2024enhancement} values.

$^i$ Number of cycles was calculated based on the reported GPC and target film thickness (160 \AA  )\cite{jeon2024synthesis} values.

$^j$  GPC calculated at the top of the structure (2 nm/730 cycles)\cite{nguyen2022gradient}.

$^k$  GPC calculated at the top of the structure (10 nm/730 cycles)\cite{nguyen2022gradient}.

* Asterisks are used as visual placeholders for the purge steps.
\end{itemize}
\end{flushleft}

\end{sidewaystable}

\newpage

\section{Terminology}

\begin{figure}
    \centering
    \includegraphics[width=0.4\linewidth]{Supplementary_info/esi_figures/cubeMZ_x_2026-09-03.jpg}
    \caption{A cube-cell model for average monolayer characteristics of material $\mathrm{MZ}_x$, with average monolayer height $\bar{h}_{\mathrm{ml}}$ (nm), average surface area per  MZ$_x$ unit $\bar{a}_{\mathrm{MZ_x}}$ (nm$^2$), and average areal number density of metal M atoms (\#/nm$^2$) (updated from Ref.~\cite{puurunen2003growth}).  }
    \label{fig:cube}
\end{figure}

In this work,  \textit{partial conformality analysis} is performed by studying  ALD growth into lateral high-aspect-ratio (LHAR) structures that are square-shaped and have a defined channel length $L$, height $H$ and width $W$. Studies into such LHAR structures can be made both through simulations and experiments; for experiments the PillarHall\textsuperscript{TM} test structures are often used\cite{cremers2019conformality,yim2020saturation,haimi2026atomic,philip2024conformal,van2023excellent,philip20233d} and are employed in this work. A schematic of a mirrored PillarHall\textsuperscript{TM} LHAR structure  is shown in Figure~\ref{fig:pillarhallscheme}a). When the exposures of reactants (exposure is expressed as partial pressure $\times$ exposure time; the absolute number of molecules delivered depends also on temperature) is such that the film does not penetrate to the end of the LHAR structure, the film is \textit{partially conformal} and a characteristic \textit{thickness profile} forms (Figure~\ref{fig:pillarhallscheme}b). In PillarHall\textsuperscript{TM} test structures, the top membrane is removed to expose the thickness profile for detailed analysis (Figure~\ref{fig:pillarhallscheme}c). 

\begin{figure}[H]
    \centering
    \includegraphics[width=1\linewidth]{Supplementary_info/FigureS2_pillarhallscheme.jpg}
    \caption{ Schematic representation of an ALD film in a mirrored lateral high-aspect-ratio structure (example: PillarHall$^{\mathrm{TM}}$), with a channel height $H$, length $L$ and width $W$ indicated. Top and side views are shown. Test structures shown are: (a) before ALD, (b) after ALD, and (c) after peeling the top membrane.}
\label{fig:pillarhallscheme}
\end{figure}

The resulting characteristic thickness profile, also sometimes in ALD called the \textit{saturation profile}, examples of which are shown in Figure~\ref{fig:schematicprofile}, can be presented in many ways. For terminology, we follow the classification by Yim et al. \cite{yim2020saturation,yim2022conformality}: the \textit{as-measured thickness/intensity/saturation profile} (with the distance $x$ into the LHAR channel as abscissa and the original measurement as ordinate); the \textit{scaled thickness/intensity/saturation profile} (with the dimensionless distance into the LHAR channel, given by $x$ divided by channel height $H$, as abscissa and the original measurement divided by the number of ALD cycles $N$ as ordinate); Type 1 normalized thickness/intensity profile (with the relative distance into the LHAR channel $x/H$ as abscissa and the normalized original measurement as ordinate; normalization is done to the beginning of the LHAR channel); and Type 2 normalized thickness/intensity profile (with the normalized distance into the LHAR channel $x/L$ where L is LHAR channel length as abscissa and the normalized original measurement as ordinate). The partial conformality analysis of this work employs original intensity/saturation profiles and Type 1 normalized saturation profiles.

\begin{figure}[H]
    \centering
    \includegraphics[width=1\linewidth]{Maindoc/figures/zonesexplain_2026-08-14.jpg}
    \caption{Scheme for partial conformality analysis showing examples of simulated (a) and measured (b) saturation/thickness/intensity profiles with classification (similar as in Ref. \cite{yim2020saturation}) into distinct \textit{zones} indicated. Zone I: Located in front of the LHAR channel. In the mirrored test structures, Zone I is located at the center. Zone II: Begins at the channel entrance marked ($x=0$), on either sides of Zone I in the mirrored test structures. Zone II extends into the channel until the intensity starts to strongly decrease. Zone III: characterized by strong decrease in the film thickness, representing the \textit{adsorption front}. Of special interest related to the analysis of adsorption kinetics, is the slope of the adsorption front. Zone IV: Located towards the end of the LHAR channel where the film thickness is close to zero. (At the very end, close to point $x=L$, a trunk of increasing signal may rise, which would be Zone V, not indicated in the scheme.) }
    \label{fig:schematicprofile}
\end{figure}

Detailed partial conformality analysis of ALD growth benefits from dividing the thickness/\-intensity/saturation profiles further into distinct parts. We again follow the classification developed by Yim et al. \cite{yim2020saturation} (however, we chose to refer to \textit{zones} instead of regions). The zones are illustrated in Figure~\ref{fig:schematicprofile}. Zone I is outside of the LHAR channel; in the PillarHall\textsuperscript{TM} test structures this represents the open area in front of the channel, where film thickness presumably is  similar as on flat silicon wafers. Zone II starts where the LHAR channel starts (distance $x=0$), and continues until the point where thickness starts to sharply decreases, at a "knee point". In simulations based on ideal ALD, the thickness/coverage in Zone II is constant (Figure~\ref{fig:schematicprofile}a), while experimental profiles may show various non-ideal features (Figure~\ref{fig:schematicprofile}b) that reveal information of the ALD process details. Zone II is further divided into regions IIa, IIb and IIc, where IIa marks the very beginning of the LHAR channel (here, intensity can increase as in the example of panel b or also decrease as observed in the literature\cite{haimi2026atomic}, IIb comes thereafter and has typically a constant thickness/intensity, which is followed by IIc where intensity may decrease. Zone III starts from the knee and represents the \textit{adsorption front} where the film thickness/signal intensity sharply decreases. Zone III ends where the thickness/signal gets close to zero, and Zone IV starts, where typically the film thickness is zero. (Sometimes, in closed channels, a "trunk" of increased growth may be observed at the end \cite{jarvilehto_simulation_2023} which would represent Zone V, not shown in the scheme.) 

For analysis of kinetics of adsorption,  the adsorption front (i.e. Zone III), is especially informative. Arts et al.\cite{arts_sticking_2019} developed a method applicable to Type 1 normalized saturation profile and \textit{Knudsen diffusion} conditions to back-extract the (lumped) sticking coefficient of (irreversible single-site Langmuir) adsorption from the slope of the adsorption front (Zone III) where the normalized intensity has halved $c = 13.9 \left( \left| \frac{\mathrm{d}\theta}{\mathrm{d}(x/H)} \right|_{\theta=0.5} \right)^2$; we have employed this \textit{slope method} in this work (results presented in the supplementary information only).  

\section{Experimental}

\subsection{Simulation parameters}

\begin{table}[H]
\centering
\setlength{\tabcolsep}{3.5pt}
\small
\caption{Parameters used in the simulations$^a$}
\begin{tabular}{|C{0.12\textwidth}|C{0.17\textwidth}|C{0.09\textwidth}|C{0.12\textwidth}|C{0.12\textwidth}|C{0.13\textwidth}|C{0.09\textwidth}|C{0.12\textwidth}|}
\hline
\makecell{Channel\\height $H$\\(m)} &
\makecell{Partial pressure\\of the reactant $p_{A0}$\\(Pa)} &
\makecell{Exposure\\time $t$\\(s)} &
\makecell{Exposure\\$(p_{A0}t)$\\(Pa$\cdot$s)} &
\makecell{Sticking\\coefficient $c$\\(--)} &
\makecell{Adsorption\\capacity $q_0$\\(\#/nm$^{2}$)} &
\makecell{Channel\\width $W$\\(m)} &
\makecell{Kn number\\(calculated)\\(--)} \\
\hline
$10^{-8}$ & 0.1   & 100  & 10       & 0.001, 0.01, 1 & 0.5, 1, 2, 4, 8 & 0.01 & $2.37\times10^{5}$ \\
$10^{-8}$ & 1     & 10   & 10       & 0.001, 0.01, 1 & 0.5, 1, 2, 4, 8 & 0.01 & $2.37\times10^{4}$ \\
$10^{-8}$ & 10    & 1    & 10       & 0.001, 0.01, 1 & 0.5, 1, 2, 4, 8 & 0.01 & $2.37\times10^{3}$ \\
$10^{-8}$ & 100   & 0.1  & 10       & 0.001, 0.01, 1 & 0.5, 1, 2, 4, 8 & 0.01 & $2.37\times10^{2}$ \\
$10^{-8}$ & 1000  & 0.01 & 10       & 0.001, 0.01, 1 & 0.5, 1, 2, 4, 8 & 0.01 & $2.37\times10^{1}$ \\
$10^{-7}$ & 1000  & 0.012& 12       & 0.001, 0.01, 1 & 0.5, 1, 2, 4, 8 & 0.01 & $2.37\times10^{0}$ \\
$10^{-6}$ & 1000  & 0.035& 35       & 0.001, 0.01, 1 & 0.5, 1, 2, 4, 8 & 0.01 & $2.37\times10^{-1}$ \\
$10^{-6}$ & 10000 & 0.034& 340      & 0.001, 0.01, 1 & 0.5, 1, 2, 4, 8 & 0.01 & $2.37\times10^{-2}$ \\
$10^{-5}$ & 10000 & 0.8  & 8000     & 0.001, 0.01, 1 & 0.5, 1, 2, 4, 8 & 0.01 & $2.37\times10^{-3}$ \\
$10^{-4}$ & 10000 & 40   & 400000   & 0.001, 0.01, 1 & 0.5, 1, 2, 4, 8 & 0.1$^b$  & $2.37\times10^{-4}$ \\
$10^{-3}$ & 10000 & 2600 & 26000000  & 0.001, 0.01, 1 & 0.5, 1, 2, 4, 8 & 1$^b$    & $2.37\times10^{-5}$ \\
\hline
\end{tabular}
\begin{tablenotes}
\item[$^a$] $^a$ Other parameters used: The partial pressure of the inert gas I, $p_I$ was set to nine times the value of the partial pressure of the reactant gas $p_{A0}$ i.e. $p_I$ = $9 \times p_{A0} $. The aspect ratio ($L$/$H$) = 1000. Desorption probability $P_d$ = 0.0001 s$^{-1}$. Temperature = 573.15 K,  hard-sphere diameter of molecule A $d_A$ =  $6\times10^{-10}$ m;
hard-sphere diameter of an inert gas molecule $d_I$ =  $4\times10^{-10}$ m; molar mass of reactant A $M_A$ = 0.1 kg/mol, molar mass of the inert gas
$M_I$ = 0.028 kg/mol. 
\item[$^b$] $^b$ Channel width increased from 0.01 to ensure that $W$ $\gg$ $H$; i.e. $W$/$H$ $\geq$ 1000.
\end{tablenotes}
\label{tab:sim_parameters}
\end{table}

\section{Results}

\subsection{Simulated saturation and pressure profiles}
Figure \ref{fig:saturationprofiles} shows saturation profile results for three Kn numbers representing different diffusion regimes, and sticking coefficients. The corresponding pressure profiles are shown in Figure \ref{fig:pressureprofiles}. Exposure was varied to get comparable values of penetration depths. In all diffusion regimes, when the adsorption capacity $q_0$ decreases, the film penetration depth increases. This observation is valid for all sticking coefficients. The sticking coefficient merely affects the steepness of the adsorption front: the shape of the saturation profile transforms from step-like (steeper slope) to having softer features (less-steep slope)  when the sticking coefficient decreases. 

\begin{figure}[H]
    \centering
    \includegraphics[width=0.98\linewidth]{Maindoc/figures/saturationprofiles_2026-05-21.jpg}
    \caption{Saturation profiles in wide lateral high-aspect-ratio structures with varied adsorption capacity $q_0$ (\#/nm$^{2}$). Saturation profiles are for different sticking coefficients and Kn numbers: (a) $c$ = 0.1, Kn = 2.37 $\times 10^3$,  exposure = 10 Pa$\cdot$s; (b) $c$ = 0.1, Kn = 2.37, exposure = 12 Pa$\cdot$s;  (c) $c$ = 0.1, Kn = 2.37 $\times 10^{-3}$, exposure = 8000 Pa$\cdot$s; (d) $c$ = 0.01, Kn = 2.37 $\times 10^3$, exposure = 10 Pa$\cdot$s;  (e) $c$ = 0.01, Kn = 2.37, exposure = 12 Pa$\cdot$s;   (f) $c$ = 0.01, Kn = 2.37 $\times 10^{-3}$, exposure = 8000 Pa$\cdot$s; (g) $c$ = 0.001, Kn = 2.37 $\times 10^3$, exposure = 10 Pa$\cdot$s; (h) $c$ = 0.001, Kn = 2.37, exposure = 12 Pa$\cdot$s; and (i) $c$ = 0.001, Kn = 2.37 $\times 10^{-3}$, exposure = 8000 Pa$\cdot$s. Corresponding pressure profiles are in Figure \ref{fig:pressureprofiles}. Extracted penetration depth and slope values are shown as Tables \ref{tab:c0.001,theta=0.5}, \ref{tab:c0.01,theta=0.5}, and \ref{tab:c0.1,theta=0.5}. Detailed simulation parameters are in Table \ref{tab:sim_parameters}.}
    \label{fig:saturationprofiles}
\end{figure}

\begin{figure}[H]
    \centering
    \includegraphics[width=0.98\linewidth]{Supplementary_info/esi_figures/pressureprofiles_2026-06-23.jpg}
    \caption{Pressure profiles corresponding to the saturation profiles of Figure \ref{fig:saturationprofiles}. The pressure profiles are for different sticking coefficients and Kn numbers: (a) $c$ = 0.1, Kn = 2.37 $\times 10^3$; (b) $c$ = 0.1, Kn = 2.37, exposure = 12 Pa$\cdot$s;  exposure = 10 Pa$\cdot$s; (c) $c$ = 0.1, Kn = 2.37 $\times 10^{-3}$, exposure = 8000 Pa$\cdot$s; (d) $c$ = 0.01, Kn = 2.37 $\times 10^3$, exposure = 10 Pa$\cdot$s;  (e) $c$ = 0.01, Kn = 2.37, exposure = 12 Pa$\cdot$s;   (f) $c$ = 0.01, Kn = 2.37 $\times 10^{-3}$, exposure = 8000 Pa$\cdot$s; (g) $c$ = 0.001, Kn = 2.37 $\times 10^3$, exposure = 10 Pa$\cdot$s; (h) $c$ = 0.001, Kn = 2.37, exposure = 12 Pa$\cdot$s; and (i) $c$ = 0.001, Kn = 2.37 $\times 10^{-3}$, exposure = 8000 Pa$\cdot$s. Detailed simulation parameters are in Table \ref{tab:sim_parameters}.} 
    \label{fig:pressureprofiles}
\end{figure}

 \begin{figure}[H]
    \centering
    \includegraphics[width=0.7\linewidth]{Maindoc/figures/simualtioncombine_2026-08-24.jpg}
    \caption{(a) Penetration depth at half coverage $(x/H)_{\theta =0.5}$ (-) as a function of the adsorption capacity $q_0$  for a range of Knudsen numbers. The sticking coefficient was $c$ = 0.01. (b) Relative penetration depth as a function of the adsorption capacity $q_0$  for a range of Knudsen numbers. The value of penetration depth at $q_0$ = 4 \#/nm$^{2}$ was taken as a reference. The background lines represent the generalized power-law fit of the form $y = Cx^b$. The sticking coefficient $c$ was 0.01. Knudsen number ranges are shown, exact values are as in the legend of Figure \ref{fig:inverse}a.} 
    \label{fig:inverse}
\end{figure}

  \begin{figure}[H]
    \centering
    \includegraphics[width=1\linewidth]{Supplementary_info/esi_figures/diffc_profiles_2026-05-29.jpg}
    \caption{Saturation profiles with varied sticking coefficients representing diffusion regimes with Kn numbers: (a) 2.37 $\times$ 10$^3$, (b) 2.37, and (c) 2.37 $\times$ 10$^{-3}$. The saturation profiles cross-over each other at a value of surface coverage ($\theta$) of 0.42. The saturation profiles are from Figure \ref{fig:saturationprofiles} for an adsorption capacity $q_0$ of 4 \#/nm$^2$.}
    \label{fig:diffc_pivotpoint}
\end{figure}

\subsection{Power-law fitting examples}

 \begin{figure}[H]
    \centering
    \includegraphics[width=1\linewidth]{Supplementary_info/esi_figures/FigS8_representative_powerlaw_fits.png}
    \caption{Representative power-law fit of penetration depth at half coverage ($(x/H)_{\theta =0.5}$) as a function of adsorption capacity $q_0$ for different diffusion regimes with Kn numbers: (a) 2.37 $\times$ 10$^3$, (b) 2.37, and (c) 2.37 $\times$ 10$^{-3}$. The sticking coefficient was $c$ = 0.01. In all panels, the dotted line represents a nonlinear power-law regression of the form $y = Cx^b$.}
    \label{fig:fitting_examples}
\end{figure}

\subsection{Power-Law exponents}

To visualize the strength of the correlation of penetration depth on the adsorption capacity across different diffusion regimes, the exponent $b$ of the fitting equation $(x/H)_{\theta=0.5}$ = $C\,q_{0}^{b}$ is plotted in Figure \ref{fig:exponents_allc}a. In the Knudsen diffusion regime (Kn $\gg$1, gray background), the value of the exponent is -0.5 and increases to about -0.2 in the molecular diffusion regime (Kn $\ll$1). As observed previously, and in line with the inverse correlation, for all diffusion regimes, the value of the exponent stays negative.

\begin{figure} [H]
    \centering
\includegraphics[width=\linewidth]{Maindoc/figures/exponents_2026-07-24.jpg}
    \caption{Value of exponents \textit{b} of the fitting equation ($x/H$)$_{\theta=0.5}$ = $C\,q_{0}^{b}$ of the penetration depth ($x/H$)$_{\theta=0.5}$ vs. adsorption capacity $q_0$ for different Kn numbers: (a) penetration depth extracted at half surface coverage $(x/H)_{\theta =0.5}$, and (b) penetration depth extracted at surface coverage of 0.42 $(x/H)_{\theta =0.42}$. The value of the exponent remains approximately the same irrespective of the sticking coefficient. }
    \label{fig:exponents_allc}
\end{figure}

  Changing the sticking coefficient almost does not affect the relation between the penetration depth and the adsorption capacity (Figure \ref{fig:exponents_allc}). 
 In addition to the molecular diffusion and transition region, in the Knudsen diffusion regime ($Kn \gg 1$) seen from Figure \ref{fig:exponents_allc}a, the exponent values appear to decrease slightly as the sticking coefficient decreases. This is an artifact of the way the penetration depth analysis was done; penetration depth was extracted at half surface coverage (surface coverage, $\theta=0.5$) of the Type-I normalized thickness profile as done in the literature\cite{yim2020saturation,yim2022conformality,jarvilehto_simulation_2023,heikkinen2024atomic,gonsalves2024simulated}. If the penetration depth would be extracted exactly at the point where the saturation profiles simulated with different sticking coefficients cross-over each other (Figure \ref{fig:diffc_pivotpoint}), in this case at $\theta=0.42$, the penetration depths would be similar irrespective of the sticking coefficient. Figure \ref{fig:exponents_allc}b presents the exponent analysis done with penetration depths extracted at $\theta=0.42$. Across all diffusion regimes, including the Knudsen diffusion regime (Kn $\gg$ 1), the inverse square root relationship between penetration depth and adsorption capacity is independent of changes in the sticking coefficient.

\clearpage
\subsection{Extracted simulated parameters }

\begin{table}[ht]
\centering
\small
\caption{Penetration depth [$(x/H)_{\theta=0.5}$] and absolute value of the slope for different adsorption capacities $q_0$ and Knudsen numbers extracted from simulated saturation profiles. The sticking coefficient $c$ was 0.001. Penetration depth and slope were analyzed at half surface coverage ($\theta=0.5$). Values of penetration depth were used to extract the power-law exponent $b$ in Figure \ref{fig:exponents_allc}a. }
\begin{tabular}{|l|l|r|r|r|r|r|}
\hline
\multicolumn{2}{|c|}{} & \multicolumn{5}{c|}{Adsorption capacity $q_0$\,(\#/nm$^{2}$)} \\
Knudsen number (-) & Parameter & 0.5 & 1 & 2 & 4 & 8 \\
\hline
$2.37\times 10^{5}$ & $(x/H)_{\theta=0.5}$ & 758.51 & 533.41 & 373.68 & 259.98 & 178.54 \\
                    & $\lvert\text{slope}\rvert$         &   0.0085 &   0.0085 &   0.0085 &   0.0085 &   0.0085 \\
\hline
$2.37\times 10^{4}$ & $(x/H)_{\theta=0.5}$ & 758.46 & 533.33 & 373.61 & 259.95 & 178.52 \\
                    & $\lvert\text{slope}\rvert$         &   0.0085 &   0.0085 &   0.0085 &   0.0085 &   0.0085 \\
\hline
$2.37\times 10^{3}$ & $(x/H)_{\theta=0.5}$ & 758.39 & 533.30 & 373.61 & 259.93 & 178.50 \\
                    & $\lvert\text{slope}\rvert$         &   0.0085 &   0.0085 &   0.0085 &   0.0085 &   0.0085 \\
\hline
$2.37\times 10^{2}$ & $(x/H)_{\theta=0.5}$ & 757.53 & 532.71 & 373.22 & 259.61 & 178.31 \\
                    & $\lvert\text{slope}\rvert$         &   0.0085 &   0.0085 &   0.0085 &   0.0100 &   0.0085 \\
\hline
$2.37\times 10^{1}$ & $(x/H)_{\theta=0.5}$ & 749.43 & 527.03 & 369.26 & 256.89 & 176.42 \\
                    & $\lvert\text{slope}\rvert$         &   0.0086 &   0.0086 &   0.0086 &   0.0086 &   0.0086 \\
\hline
$2.37\times 10^{0}$ & $(x/H)_{\theta=0.5}$ & 745.61 & 525.23 & 368.61 & 257.08 & 177.34 \\
                    & $\lvert\text{slope}\rvert$         &   0.0095 &   0.0095 &   0.0095 &   0.0095 &   0.0095 \\
\hline
$2.37\times 10^{-1}$ & $(x/H)_{\theta=0.5}$ & 758.95 & 540.77 & 382.91 & 269.82 & 189.09 \\
                     & $\lvert\text{slope}\rvert$        &   0.0157 &   0.0157 &   0.0157 &   0.0157 &   0.0157 \\
\hline
$2.37\times 10^{-2}$ & $(x/H)_{\theta=0.5}$ & 762.95 & 577.42 & 425.15 & 307.23 & 219.47 \\
                     & $\lvert\text{slope}\rvert$         &   0.0424 &   0.0424 &   0.0427 &   0.0423 &   0.0426 \\
\hline
$2.37\times 10^{-3}$ & $(x/H)_{\theta=0.5}$ & 735.18 & 626.38 & 516.03 & 409.47 & 313.26 \\
                     & $\lvert\text{slope}\rvert$         &   0.1328 &   0.1298 &   0.1322 &   0.1320 &   0.1300 \\
\hline
$2.37\times 10^{-4}$ & $(x/H)_{\theta=0.5}$ & 750.78 & 686.28 & 617.78 & 545.18 & 468.98 \\
                     & $\lvert\text{slope}\rvert$         &   0.4098 &   0.4118 &   0.4136 &   0.4045 &   0.4132 \\
\hline
$2.37\times 10^{-5}$ & $(x/H)_{\theta=0.5}$ & 758.88 & 715.42 & 669.76 & 621.57 & 570.55 \\
                     & $\lvert\text{slope}\rvert$         &   1.2496 &   1.2235 &   1.1816 &   1.2486 &   1.2420 \\
\hline
\end{tabular}
\label{tab:c0.001,theta=0.5}
\end{table}

\begin{table}[ht]
\centering
\small
\caption{Penetration depth [$(x/H)_{\theta=0.42}$] and absolute value of the slope for different adsorption capacities $q_0$ and Knudsen numbers extracted from simulated saturation profiles. The sticking coefficient $c$ was 0.001. Penetration depth and slope were analyzed at surface coverage $\theta=0.42$. Values of penetration depth were used to extract the power-law exponent $b$ of Figure \ref{fig:exponents_allc}b. }
\begin{tabular}{|l|l|r|r|r|r|r|}
\hline
\multicolumn{2}{|c|}{} & \multicolumn{5}{c|}{Adsorption capacity $q_0$\,(\#/nm$^{2}$)} \\
Knudsen number (-) & Parameter & 0.5 & 1 & 2 & 4 & 8 \\
\hline
$2.37\times 10^{5}$ & $(x/H)_{\theta=0.42}$ & 768.22 & 543.11 & 383.39 & 269.73 & 188.25 \\
                    & $\lvert\text{slope}\rvert$          &   0.0079 & 0.0079 & 0.0079 & 0.0079 & 0.0079 \\
\hline
$2.37\times 10^{4}$ & $(x/H)_{\theta=0.42}$ & 768.15 & 543.05 & 383.32 & 269.66 & 188.21 \\
                    & $\lvert\text{slope}\rvert$          &   0.0079 & 0.0079 & 0.0079 & 0.0079 & 0.0079 \\
\hline
$2.37\times 10^{3}$ & $(x/H)_{\theta=0.42}$ & 768.07 & 542.98 & 383.31 & 269.61 & 188.19 \\
                    & $\lvert\text{slope}\rvert$          &   0.0079 & 0.0079 & 0.0079 & 0.0079 & 0.0079 \\
\hline
$2.37\times 10^{2}$ & $(x/H)_{\theta=0.42}$ & 767.22 & 542.42 & 382.90 & 269.33 & 187.99 \\
                    & $\lvert\text{slope}\rvert$          &   0.0079 & 0.0079 & 0.0079 & 0.0079 & 0.0080 \\
\hline
$2.37\times 10^{1}$ & $(x/H)_{\theta=0.42}$ & 759.00 & 536.63 & 378.82 & 266.49 & 186.02 \\
                    & $\lvert\text{slope}\rvert$          &   0.0080 & 0.0080 & 0.0080 & 0.0080 & 0.0080 \\
\hline
$2.37\times 10^{0}$ & $(x/H)_{\theta=0.42}$ & 754.33 & 533.94 & 377.32 & 265.80 & 186.05 \\
                    & $\lvert\text{slope}\rvert$          &   0.0088 & 0.0088 & 0.0088 & 0.0088 & 0.0088 \\
\hline
\end{tabular}
\label{tab:c0.001,theta=0.42}
\end{table}

\begin{table}[ht]
\centering
\small
\caption{Penetration depth [$(x/H)_{\theta=0.5}$] and absolute value of the slope for different adsorption capacities $q_0$ and Knudsen numbers extracted from simulated saturation profiles. The sticking coefficient $c$ was 0.01. Penetration depth and slope were analyzed at surface coverage $\theta=0.5$. The penetration depth values are plotted as Figure 2b in the main manuscript. Also, values of penetration depth were used to extract the power-law exponent $b$ in Figure \ref{fig:exponents_allc}a.}
\begin{tabular}{|l|l|r|r|r|r|r|}
\hline
\multicolumn{2}{|c|}{} & \multicolumn{5}{c|}{Adsorption capacity $q_0$\,(\#/nm$^{2}$)} \\
Knudsen number (-) & Parameter & 0.5 & 1 & 2 & 4 & 8 \\
\hline
$2.37\times 10^{5}$ & $(x/H)_{\theta=0.5}$ & 764.06 & 539.60 & 380.83 & 268.51 & 188.96 \\
                    & $\lvert\text{slope}\rvert$         &   0.0268 &   0.0269 &   0.0268 &   0.0269 &   0.0269 \\
\hline
$2.37\times 10^{4}$ & $(x/H)_{\theta=0.5}$ & 764.06 & 539.58 & 380.83 & 268.49 & 188.96 \\
                    & $\lvert\text{slope}\rvert$         &   0.0269 &   0.0269 &   0.0268 &   0.0269 &   0.0270 \\
\hline
$2.37\times 10^{3}$ & $(x/H)_{\theta=0.5}$ & 763.96 & 539.53 & 381.00 & 268.45 & 188.94 \\
                    & $\lvert\text{slope}\rvert$         &   0.0268 &   0.0269 &   0.0268 &   0.0269 &   0.0270 \\
\hline
$2.37\times 10^{2}$ & $(x/H)_{\theta=0.5}$ & 763.13 & 538.94 & 380.37 & 268.16 & 188.73 \\
                    & $\lvert\text{slope}\rvert$         &   0.0268 &   0.0269 &   0.0271 &   0.0271 &   0.0270 \\
\hline
$2.37\times 10^{1}$ & $(x/H)_{\theta=0.5}$ & 754.93 & 533.21 & 376.34 & 265.34 & 186.74 \\
                    & $\lvert\text{slope}\rvert$         &   0.0271 &   0.0271 &   0.0272 &   0.0273 &   0.0272 \\
\hline
$2.37\times 10^{0}$ & $(x/H)_{\theta=0.5}$ & 750.53 & 530.68 & 374.80 & 264.40 & 186.20 \\
                    & $\lvert\text{slope}\rvert$         &   0.0300 &   0.0299 &   0.0298 &   0.0300 &   0.0300 \\
\hline
$2.37\times 10^{-1}$ & $(x/H)_{\theta=0.5}$ & 761.67 & 543.64 & 386.04 & 273.31 & 193.11 \\
                     & $\lvert\text{slope}\rvert$         &   0.0496 &   0.0497 &   0.0497 &   0.0494 &   0.0491 \\
\hline
$2.37\times 10^{-2}$ & $(x/H)_{\theta=0.5}$ & 763.87 & 578.35 & 426.09 & 308.22 & 220.52 \\
                     & $\lvert\text{slope}\rvert$         &   0.1360 &   0.1316 &   0.1352 &   0.1316 &   0.1354 \\
\hline
$2.37\times 10^{-3}$ & $(x/H)_{\theta=0.5}$ & 735.47 & 626.66 & 516.32 & 409.76 & 313.55 \\
                     & $\lvert\text{slope}\rvert$         &   0.4073 &   0.4127 &   0.4132 &   0.4084 &   0.4221 \\
\hline
$2.37\times 10^{-4}$ & $(x/H)_{\theta=0.5}$ & 750.86 & 686.36 & 617.86 & 545.27 & 469.07 \\
                     & $\lvert\text{slope}\rvert$         &   1.2531 &   1.2563 &   1.2452 &   1.2349 &   1.2493 \\
\hline
$2.37\times 10^{-5}$ & $(x/H)_{\theta=0.5}$ & 758.95 & 715.51 & 669.77 & 621.59 & 570.57 \\
                     & $\lvert\text{slope}\rvert$         &   1.1176 &   1.3554 &   1.1597 &   1.7485 &   1.7483 \\
\hline
\end{tabular}
\label{tab:c0.01,theta=0.5}
\end{table}

\begin{table}[ht]
\centering
\small
\caption{Penetration depth [$(x/H)_{\theta=0.42}$] and absolute value of the slope for different adsorption capacities $q_0$ and Knudsen numbers extracted from simulated saturation profiles. The sticking coefficient $c$ was 0.01. Penetration depth and slope were analyzed at surface coverage $\theta=0.42$. Values of penetration depth were used to extract the power-law exponent $b$ in Figure \ref{fig:exponents_allc}b. }
\begin{tabular}{|l|l|r|r|r|r|r|}
\hline
\multicolumn{2}{|c|}{} & \multicolumn{5}{c|}{Adsorption capacity $q_0$\,(\#/nm$^{2}$)} \\
Knudsen number (-) & Parameter & 0.5 & 1 & 2 & 4 & 8 \\
\hline
$2.37\times 10^{5}$ & $(x/H)_{\theta=0.42}$ & 767.15 & 542.67 & 383.92 & 271.59 & 192.04 \\
                    & $\lvert\text{slope}\rvert$          &   0.0250 &   0.0251 &   0.0250 &   0.0252 &   0.0251 \\
\hline
$2.37\times 10^{4}$ & $(x/H)_{\theta=0.42}$ & 767.13 & 542.67 & 383.91 & 271.57 & 192.03 \\
                    & $\lvert\text{slope}\rvert$          &   0.0250 &   0.0251 &   0.0250 &   0.0252 &   0.0252 \\
\hline
$2.37\times 10^{3}$ & $(x/H)_{\theta=0.42}$ & 767.04 & 542.62 & 383.86 & 271.53 & 192.01 \\
                    & $\lvert\text{slope}\rvert$          &   0.0250 &   0.0251 &   0.0250 &   0.0252 &   0.0252 \\
\hline
$2.37\times 10^{2}$ & $(x/H)_{\theta=0.42}$ & 766.19 & 542.02 & 383.44 & 271.24 & 191.80 \\
                    & $\lvert\text{slope}\rvert$          &   0.0251 &   0.0251 &   0.0251 &   0.0250 &   0.0251 \\
\hline
$2.37\times 10^{1}$ & $(x/H)_{\theta=0.42}$ & 757.97 & 536.25 & 379.38 & 268.36 & 189.77 \\
                    & $\lvert\text{slope}\rvert$          &   0.0252 &   0.0255 &   0.0253 &   0.0253 &   0.0253 \\
\hline
$2.37\times 10^{0}$ & $(x/H)_{\theta=0.42}$ & 753.29 & 533.43 & 377.55 & 267.16 & 188.96 \\
                    & $\lvert\text{slope}\rvert$          &   0.0280 &   0.0278 &   0.0280 &   0.0277 &   0.0280 \\
\hline
\end{tabular}
\label{tab:c0.01,theta=0.42}
\end{table}

\begin{table}[ht]
\centering
\small
\caption{Penetration depth [$(x/H)_{\theta=0.5}$] and absolute value of the slope for different adsorption capacities $q_0$ and Knudsen numbers at surface coverage $\theta=0.5$. The sticking coefficient $c$ was 0.1. Values of penetration depth were used to extract the power-law exponent $b$ in Figure \ref{fig:exponents_allc}a. }
\begin{tabular}{|l|l|r|r|r|r|r|}
\hline
\multicolumn{2}{|c|}{} & \multicolumn{5}{c|}{Adsorption capacity $q_0$\,(\#/nm$^{2}$)} \\
Knudsen number (-) & Parameter & 0.5 & 1 & 2 & 4 & 8 \\
\hline
$2.37\times 10^{5}$ & $(x/H)_{\theta=0.5}$ & 765.50 & 541.11 & 382.43 & 270.00 & 190.86 \\
                    & $\lvert\text{slope}\rvert$         &   0.0841 &   0.0847 &   0.0849 &   0.0838 &   0.0845 \\
\hline
$2.37\times 10^{4}$ & $(x/H)_{\theta=0.5}$ & 765.50 & 541.10 & 382.43 & 270.21 & 190.86 \\
                    & $\lvert\text{slope}\rvert$         &   0.0843 &   0.0852 &   0.0848 &   0.0840 &   0.0850 \\
\hline
$2.37\times 10^{3}$ & $(x/H)_{\theta=0.5}$ & 765.41 & 541.04 & 382.39 & 270.18 & 190.84 \\
                    & $\lvert\text{slope}\rvert$         &   0.0838 &   0.0848 &   0.0847 &   0.0840 &   0.0849 \\
\hline
$2.37\times 10^{2}$ & $(x/H)_{\theta=0.5}$ & 764.57 & 540.45 & 381.97 & 269.90 & 190.63 \\
                    & $\lvert\text{slope}\rvert$         &   0.0846 &   0.0846 &   0.0851 &   0.0850 &   0.0840 \\
\hline
$2.37\times 10^{1}$ & $(x/H)_{\theta=0.5}$ & 756.36 & 534.70 & 377.92 & 267.03 & 188.62 \\
                    & $\lvert\text{slope}\rvert$         &   0.0856 &   0.0860 &   0.0857 &   0.0865 &   0.0862 \\
\hline
$2.37\times 10^{0}$ & $(x/H)_{\theta=0.5}$ & 751.82 & 532.01 & 376.21 & 265.92 & 187.87 \\
                    & $\lvert\text{slope}\rvert$         &   0.0954 &   0.0944 &   0.0931 &   0.0946 &   0.0946 \\
\hline
$2.37\times 10^{-1}$ & $(x/H)_{\theta=0.5}$ & 762.41 & 544.41 & 386.83 & 274.14 & 194.00 \\
                     & $\lvert\text{slope}\rvert$         &   0.1520 &   0.1541 &   0.1539 &   0.1517 &   0.1578 \\
\hline
$2.37\times 10^{-2}$ & $(x/H)_{\theta=0.5}$ & 764.14 & 578.62 & 426.37 & 308.50 & 220.81 \\
                     & $\lvert\text{slope}\rvert$         &   0.4159 &   0.4224 &   0.4029 &   0.4232 &   0.4086 \\
\hline
$2.37\times 10^{-3}$ & $(x/H)_{\theta=0.5}$ & 735.58 & 626.73 & 516.42 & 409.87 & 313.62 \\
                     & $\lvert\text{slope}\rvert$         &   1.0348 &   1.0606 &   0.9844 &   0.9826 &   1.0042 \\
\hline
$2.37\times 10^{-4}$ & $(x/H)_{\theta=0.5}$ & 750.88 & 686.40 & 617.91 & 545.32 & 469.08 \\
                     & $\lvert\text{slope}\rvert$         &   1.6970 &   2.3649 &   1.5797 &   2.1749 &   1.9803 \\
\hline
$2.37\times 10^{-5}$ & $(x/H)_{\theta=0.5}$ & 759.00 & 715.54 & 669.80 & 621.60 & 570.57 \\
                     & $\lvert\text{slope}\rvert$         &   1.3036 &   1.5527 &   1.0852 &   1.7302 &   1.7455 \\
\hline
\end{tabular}
\label{tab:c0.1,theta=0.5}
\end{table}

\begin{table}[ht]
\centering
\small
\caption{Penetration depth [$(x/H)_{\theta=0.42}$] and absolute value of the slope for different adsorption capacities $q_0$ and Knudsen numbers at surface coverage $\theta=0.42$. The sticking coefficient $c$ was 0.1. Values of penetration depth were used to extract the power-law exponent $b$ in Figure \ref{fig:exponents_allc}b. }
\begin{tabular}{|l|l|r|r|r|r|r|}
\hline
\multicolumn{2}{|c|}{} & \multicolumn{5}{c|}{Adsorption capacity $q_0$\,(\#/nm$^{2}$)} \\
Knudsen number (-) & Parameter & 0.5 & 1 & 2 & 4 & 8 \\
\hline
$2.37\times 10^{5}$ & $(x/H)_{\theta=0.42}$ & 766.48 & 542.08 & 383.40 & 271.19 & 191.84 \\
                    & $\lvert\text{slope}\rvert$          &   0.0778 &   0.0793 &   0.0786 &   0.0774 &   0.0788 \\
\hline
$2.37\times 10^{4}$ & $(x/H)_{\theta=0.42}$ & 766.47 & 542.08 & 383.40 & 271.19 & 191.84 \\
                    & $\lvert\text{slope}\rvert$          &   0.0778 &   0.0794 &   0.0787 &   0.0775 &   0.0790 \\
\hline
$2.37\times 10^{3}$ & $(x/H)_{\theta=0.42}$ & 766.38 & 542.02 & 383.36 & 271.15 & 191.82 \\
                    & $\lvert\text{slope}\rvert$          &   0.0806 &   0.0790 &   0.0784 &   0.0772 &   0.0788 \\
\hline
$2.37\times 10^{2}$ & $(x/H)_{\theta=0.42}$ & 765.56 & 541.42 & 382.94 & 270.87 & 191.60 \\
                    & $\lvert\text{slope}\rvert$          &   0.0782 &   0.0782 &   0.0792 &   0.0790 &   0.0772 \\
\hline
$2.37\times 10^{1}$ & $(x/H)_{\theta=0.42}$ & 757.32 & 535.66 & 378.88 & 267.99 & 189.58 \\
                    & $\lvert\text{slope}\rvert$          &   0.0813 &   0.0811 &   0.0794 &   0.0807 &   0.0817 \\
\hline
$2.37\times 10^{0}$ & $(x/H)_{\theta=0.42}$ & 752.70 & 532.89 & 377.08 & 266.79 & 188.74 \\
                    & $\lvert\text{slope}\rvert$          &   0.0885 &   0.0867 &   0.0892 &   0.0870 &   0.0870 \\
\hline
\end{tabular}
\label{tab:c=0.1,theta=0.42}
\end{table}

\clearpage

\subsection{Thin film measurements}

\begin{table}[H]
\centering
\caption{Film thickness and density measurements for 180 cycles of ZnO grown at different temperatures with and without methanol.}
\label{tab:methanol_inhibitor_bordered}
\small
\setlength{\tabcolsep}{4pt} 
\begin{tabular}{|C{1.8cm}|C{0.9cm}|C{0.9cm}|C{1.4cm}|C{1.9cm}|C{2.5cm}|C{1.9cm}|C{1.9cm}|}
\hline
\thead{Sample\\code} &
\multicolumn{2}{c|}{\thead{Inhibitor\\(MeOH)}} &
\thead{ALD\\temp.\\} &
\thead{Thickness\\(Ellips.)\\} &
\thead{Thickness\\(XRR)\\} &
\thead{Rougness\\(XRR)\\} &
\thead{Film\\density\\(XRR)} \\
\hline
 & \thead{No} & \thead{Yes} & (°C)& (nm) &(nm) & (nm) & (g/cm$^3$)\\
\hline
AP251 & X &   & 150 & 29.7 & 30.2 & 2.1 & 5.3 \\

AP252 &   & X & 150 & 12.9 & 13.1 & 1.8 & 5.3 \\

AP256 & X &   & 175 & 28.4 & 29.3 & 1.4 & 5.2 \\

AP253 &   & X & 175 & 12.4 & 13.6 & 0.9 & 5.3 \\

AP255 & X &   & 200 & 25.4 & 25.7 & 1.2 & 5.0 \\

AP254 &   & X & 200 & 12.0 & 12.5 & 1.5 & 5.4 \\
\hline
\end{tabular}
\label{tab:ellipsomter_XRR}
\end{table}

\subsection{Features of experimental saturation profiles}
Saturation profiles from ALD experiments to grow zinc oxide on LHAR test chips were analyzed with SEM-EDS. 
Three repetitions were made for each SEM-EDS analysis, at different locations on the sample. Some saturation profiles have been published for the standard non-inhibited  ZnO process before \cite{haimi2026atomic} and the current results are in line with previous literature, the current series is to the authors' knowledge the most complete temperature series existing as of today for this ALD process. 
 \begin{figure}[H]
    \centering
    \includegraphics[width=1\linewidth]{Maindoc/figures/growin_2026-08-25.jpg}
    \caption{Experimental saturation profiles within the LHAR structure measured with SEM-EDS. Panels on the left side (a, c, e) show the as-measured Zn signal intensity profiles, while the panels on the right (b, d, f) show the corresponding Type I normalized saturation profiles. ALD process temperatures: (a, b) 150 $\degree$C, (c, d) 175 $\degree$C, and (e, f) 200 $\degree$C. Zone I (open area in front of the LHAR channel) is indicated with a gray background and the zero point indicates the channel entrance (beginning of Zone II). Black curves represent the standard DEZ/water process to grow ALD ZnO, while the blue curves represent the methanol-inhibited process (MeOH/DEZ/water). The number of ALD cycles was 180 in all cases.}
    \label{fig:growin_experiment}
\end{figure}

\begin{figure}[H]
    \centering
    \includegraphics[width=0.6\linewidth]{Maindoc/figures/growinexpt_2026-09-09.jpg}
    \caption{Penetration depth (extracted from the data of Figure \ref{fig:growin_experiment}) as function of the adsorption capacity $q_0$ (related to thickness-based GPC through mass density; values in Table 1 of the main manuscript and also in Table \ref{tab:signal_intensity}, Zone IIb). The dashed line shows an inverse square root relation $y=Cx^b$, where $C$ is a constant and $b=-0.5$ (here, $C=765$, from a fit with $R^2$ of 0.9973). Each experiment is represented by six repeated penetration depth values (details are in Table \ref{tab:sem-eds-sideways}).  }
    \label{fig:compare}
\end{figure}

Closer examination of the profiles of Figure \ref{fig:growin_experiment} reveals interesting differences between the saturation profile features of ZnO grown at different temperatures and without/with inhibitor. First, the detailed features of the saturation profile of the standard non-inhibited ALD ZnO process are influenced by temperature. The profile of the ZnO process at 175$^\degree$C closely resembles the profile of an ideal ALD process where the thickness is the same in Zones I and II of the profile, adsorption front is sharp in Zone III, and the signal is zero in Zone IV. The profile of the ZnO process at 200$^\degree$C resembles that of 175$^\degree$C, only a slight dip in Zn signal is observed in Zone I compared to Zone II. The profile of ZnO grown at 150$^\degree$C differs visually from the two other profiles in that the Zn signal continuously decreases with distance in Zone II, and the adsorption front in Zone III is visibly less steep. Numerical slope values are found in Table \ref{tab:slopes_stickingcoeff} and numerical Zn signal intensities are collected in Table \ref{tab:signal_intensity}.  Second, when comparing the methanol-inhibited ZnO processes to the standard non-inhibited ones, in addition to the lower GPC and deeper penetration mentioned earlier, we notice that all methanol-inhibited processes have a lower Zn signal in Zone I compared to Zone IIb (ratio ca. 0.8, see Table \ref{tab:signal_intensity}); all profiles have a decreasing Zn signal in Zone II, and all profiles show a relatively soft adsorption front in Zone III (numerical slope values are found in Table \ref{tab:slopes_stickingcoeff}. 

Arts et al.\cite{arts_sticking_2019} developed a method to extract information of adsorption kinetics directly from  a the Type I normalized saturation profile, by considering the slope of the adsorption front (i.e., Zone III), which we will apply to the current dataset. Interestingly, this "slope method" allows the direct extraction of the sticking coefficient of a reactant directly from the value of the slope, without any knowledge of the chemical nature of the reactant, provided that the conditions are valid where the slope method applies (Knudsen diffusion) \cite{arts_sticking_2019}. The slopes measured for the adsorption fronts of Figure \ref{fig:growin_experiment} are collected in Table \ref{tab:slopes_stickingcoeff}, along with the by-the-slope-method calculated sticking coefficients. For the standard DEZ/water process to grow ZnO, as the temperature increased from 150 to 200$\degree$C, the slope of the saturation profile became steeper and the extracted sticking coefficient increased from about 0.0009 to about 0.01 (Table \ref{tab:slopes_stickingcoeff}; plot as function of temperature in shown as Figure \ref{fig:growin_sticking}). The values obtained for 175-200$\degree$C are in line with previous literature values reported for the sticking coefficient of DEZ on ZnO (0.007 at 175 $\degree$C \cite{elam2003conformal} and 0.002 at 200 $\degree$C\cite{haimi2026atomic}).  
For the methanol-inhibited process, temperature had little effect, and the sticking coefficient values all remained all below 0.001, resembling that of the standard ZnO process at 150 $\degree$C. Clearly, what comes to adsorption kinetics, the methanol-inhibited process differed from the standard ZnO process.

\begin{table}[h!]
\centering
\begin{threeparttable}
\caption{Extracted slopes at half coverage and back-extracted sticking coefficients from experimental ALD saturation profiles corresponding to the standard (DEZ/H$_2$O) and the inhibited (MeOH/DEZ/H$_2$O) process.}
\label{tab:slopes_stickingcoeff}

\begin{tabular}{|C{1.8cm}|C{3cm}|C{3cm}|C{3.0cm}|C{3.0cm}|}
\hline
\multirow{2}{*}{\makecell{ALD\\temperature\\($^\circ$C)}} &
\multicolumn{2}{C{6cm}|}{\makecell{\gape Absolute value of the slope of normalized\\ intensity\tnote{a}\\ vs.\ $x/H$ (-)}} &
\multicolumn{2}{C{6.0cm}|}{\makecell{\gape Lumped sticking coefficient\tnote{a,b}\\(-)}} \\
\cline{2-5}
& DEZ/water & MeOH/DEZ/water & DEZ/water & MeOH/DEZ/water \\
\hline\hline
150 & 0.00784 $\pm 6.3\times10^{-4}$ & 0.00534 $\pm 5.4\times10^{-4}$ & $8.60\times10^{-4} \pm 1.39\times10^{-4}$ & $4.00\times10^{-4} \pm 8.06\times10^{-5}$ \\
175 & 0.01915 $\pm 5.9\times10^{-4}$ & 0.00762 $\pm 5.4\times10^{-4}$ & $5.10\times10^{-3} \pm 3.12\times10^{-4}$ & $8.11\times10^{-4} \pm 1.09\times10^{-4}$ \\
200 & 0.02972 $\pm 1.7\times10^{-3}$ & 0.00687 $\pm 1.2\times10^{-3}$ & $1.23\times10^{-2} \pm 1.40\times10^{-3}$ & $6.73\times10^{-4} \pm 2.32\times10^{-4}$ \\
\hline
\end{tabular}

\begin{tablenotes}[flushleft]\footnotesize
\item[a] Average over six line scans (see Table \ref{tab:sem-eds-sideways}).
\item[b] Calculated with the slope method \cite{arts_sticking_2019}.
\end{tablenotes}
\end{threeparttable}
\end{table}

The observed variations in the saturation profile shapes (Zone II) and adsorption front slopes (Zone III) can likely be understood by considering which reactant is limiting the growth. For the standard ZnO process at 175 and 200$\degree$C, the steeper slopes and the correspondence to previous literature\cite{haimi2026atomic} suggests that most likely, the exposure of DEZ is limiting the growth, while water is in excess. The lowest temperature of 150$\degree$C seems to differ from the rest both what comes to the shape (Zone II) and steepness of the adsorption front (Zone III). We speculate that at this lowest temperature, water starts to limit the penetration depth instead of DEZ. Somewhat analogous transition from metal-reactant-limited to water-limited penetration has been observed before for the trimethylaluminium(TMA)-water process, although there, the exposure of TMA was increased instead of decreasing temperature \cite{arts_sticking_2019, yim2020saturation}. Having water as limiting reactant, with slower and not fully saturated reactions, can be in line with the decrease of Zn intensity in Zone II, as the partial pressure of the limiting reactant decreases when progressing into the channel (Figure \ref{fig:pressureprofiles}). The fact that for all studied temperatures, the sticking coefficients measured for the methanol-inhibited process were low and at the same level as for the standard ZnO process at 150$\degree$C, may indicate that the methanol-inhibited ALD ZnO process was at water-limited conditions at all temperatures. Here, it is assumed that the small methanol molecules diffuse at least to the same depth into the HAR structure as the reactants and saturate the available surface sites. The extended film growth (higher penetration depth) compared to the standard ALD ZnO process may have led to the water-limited conditions: while the adsorption capacity of Zn had decreased through the methanol inhibitor, water had to react with all ligands (ethyl and methoxy) and would presumably be consumed along the channel rather similarly as in the non-inhibited process, hence potentially switching the limiting reactant from DEZ to water.  

It should be noted that the origin of the difference in the Zn signal intensity in Zone I vs. Zone IIa especially in the methanol-inhibited but also to a slight extent in the standard ALD ZnO process at 200$\degree$C is currently not well understood. In the literature, an analogous, but still stronger, difference has been observed for the same ALD ZnO process at 250$\degree$C. We speculate this feature to be related to desorption of some zinc-containing species either during the purges or the methanol exposure. Definite conclusion of the origin would require more experiments, preferably complemented with first principles calculations. Making such investigations is outside of the scope of this work.

\subsection{SEM-EDS results}

\begin{figure}[!ht]
    \centering
    \vspace*{\fill}
    \includegraphics[width=\linewidth, height=0.88\textheight, keepaspectratio]{Supplementary_info/esi_figures/EDScombine2026-08-07.jpg}
    \caption{Scanning electron microscopy images with elemental mapping for the (a,b,c,d): DEZ/H$_2$O process at 150 $\degree$C; (e,f,g,h): MeOH/DEZ/H$_2$O process at 150 $\degree$C; (i,j,k,l): DEZ/H$_2$O process at 175 $\degree$C; (m,n,o,p): MeOH/DEZ/H$_2$O process at 175  $\degree$C; (q,r,s,t): DEZ/H$_2$O process at 200 $\degree$C; (u,v,w,x): MeOH/DEZ/H$_2$O process at 200 $\degree$C; and MeOH/DEZ/H$_2$O process at 200 $\degree$C. The images were taken on a peeled LHAR test chip exposing the ZnO film  after 180 cycles of ALD.}
    \label{fig:SEM-EDS_combine}
    \vspace*{\fill}
   
\end{figure}

 \clearpage

\begin{sidewaystable}[p]
\captionsetup[sidewaystable]{position=top}
\centering
\caption{SEM-EDS line scans: Signal intensities in Zones I and IIb. Average values and intensity ratios are also calculated.$^a$}
\footnotesize
\setlength{\tabcolsep}{4pt}
\renewcommand{\arraystretch}{1.12}

\begin{adjustbox}{max height=0.97\textheight,keepaspectratio}
  \begin{tabular}{|M{2.15cm}|M{0.7cm}|M{1.2cm}|M{1.4cm}|M{1.9cm}|M{1.4cm}|M{2.0cm}|M{1.3cm}|M{1.6cm}|M{1.6cm}|M{2.2cm}|M{2.0cm}|M{2.0cm}|}
\hline
\makecell[l]{\rule{0pt}{3.2ex}SEM-EDS line \\scan number } &
\makecell[l]{\rule{0pt}{3.2ex}ALD\\temp.\\ (\si{\degreeCelsius})} &
\makecell[l]{\rule{0pt}{3.2ex}Inhibitor} &
\makecell[l]{\rule{0pt}{3.2ex}Zone I \\Zn signal\\ (counts)} &
\makecell[l]{\rule{0pt}{3.2ex}Avg. Zone I\\Zn signal\\(counts)$^b$} &
\makecell[l]{\rule{0pt}{3.2ex}Zone IIb \\Zn signal \\(counts)} &
\makecell[l]{\rule{0pt}{3.2ex}Avg. Zone IIb\\Zn signal\\(counts)$^b$} &
\makecell[l]{\rule{0pt}{3.2ex}Range for \\Zone IIb\\ (\si{\micro\meter})} &
\makecell[l]{\rule{0pt}{3.2ex}Background\\ Zn signal\\(counts)} &
\makecell[l]{\rule{0pt}{3.2ex}Range for\\background \\signal (\si{\micro\meter})} &
\makecell[l]{\rule{0pt}{3.2ex} Zn signal ratio: \\Zone I/\\Zone IIb$^c$} & 
\makecell[l]{\rule{0pt}{3.2ex} Zn signal ratio \\(Zone I)$^d$}& 
\makecell[l]{\rule{0pt}{3.2ex} Zn signal ratio \\(Zone IIb)$^d$}\\

\hline

251-1/left  & 150 & -    & $1452 \pm 32$ & \multirow{6}{*}{$1444 \pm 15$} & $1461 \pm 38$ & \multirow{6}{*}{$1458 \pm 17$} & 10-20 & $86 \pm 9$  & 500-700 & 0.99 & \multirow{12}{*}{0.334 } & \multirow{12}{*}{0.404} \\
251-1/right & 150 & -    & $1452 \pm 32$ &                                  & $1485 \pm 41$ &                                  & 10-20 & $91 \pm 9$  & 500-700 & 0.98 & & \\
251-2/left  & 150 & -    & $1416 \pm 38$ &                                  & $1429 \pm 40$ &                                  & 10-20 & $85 \pm 10$ & 500-700 & 0.99 & & \\
251-2/right & 150 & -    & $1416 \pm 38$ &                                  & $1429 \pm 40$ &                                  & 10-20 & $87 \pm 9$  & 500-700 & 0.99 & & \\
251-3/left  & 150 & -    & $1465 \pm 42$ &                                  & $1479 \pm 28$ &                                  & 10-20 & $87 \pm 9$  & 500-700 & 0.99 & & \\
251-3/right & 150 & -    & $1465 \pm 42$ &                                  & $1466 \pm 60$ &                                  & 10-20 & $89 \pm 9$  & 500-700 & 1.00 & & \\
\cline{1-11}

252-1/left      & 150 & MeOH & $545 \pm 27$  & \multirow{6}{*}{$543 \pm 12$} & $644 \pm 29$  & \multirow{6}{*}{$644 \pm 25$} & 20-60 & $89 \pm 9$  & 500-700 & 0.82 & & \\
252-1/right     & 150 & MeOH & $545 \pm 27$  &                                & $650 \pm 23$  &                                & 20-60 & $92 \pm 10$ & 500-700 & 0.81 & & \\
252-2/left      & 150 & MeOH & $550 \pm 32$  &                                & $646 \pm 25$  &                                & 20-60 & $90 \pm 9$  & 500-700 & 0.83 & & \\
252-2/right     & 150 & MeOH & $550 \pm 32$  &                                & $647 \pm 24$  &                                & 20-60 & $91 \pm 9$  & 500-700 & 0.83 & & \\
252-3/left      & 150 & MeOH & $536 \pm 25$  &                                & $632 \pm 25$  &                                & 20-60 & $86 \pm 9$  & 500-700 & 0.82 & & \\
252-3/right     & 150 & MeOH & $536 \pm 25$  &                                & $644 \pm 24$  &                                & 20-60 & $93 \pm 10$ & 500-700 & 0.80 & & \\
\hline

256-1/left      & 175 & -    & $1380 \pm 42$ & \multirow{6}{*}{$1354 \pm 17$} & $1413 \pm 41$ & \multirow{6}{*}{$1391 \pm 17$} & 20-60 & $93 \pm 10$ & 500-700 & 0.98 & \multirow{12}{*}{0.383 } & \multirow{12}{*}{0.475} \\
256-1/right     & 175 & -    & $1380 \pm 42$ &                                & $1427 \pm 34$ &                                & 20-60 & $95 \pm 10$ & 500-700 & 0.96 & & \\
256-2/left      & 175 & -    & $1358 \pm 43$ &                                & $1401 \pm 43$ &                                & 20-60 & $91 \pm 9$  & 500-700 & 0.97 & & \\
256-2/right     & 175 & -    & $1358 \pm 43$ &                                & $1413 \pm 39$ &                                & 20-60 & $95 \pm 11$ & 500-700 & 0.96 & & \\
256-3/left      & 175 & -    & $1325 \pm 40$ &                                & $1352 \pm 47$ &                                & 20-60 & $88 \pm 10$ & 500-700 & 0.98 & & \\
256-3/right     & 175 & -    & $1325 \pm 40$ &                                & $1341 \pm 40$ &                                & 20-60 & $90 \pm 9$  & 500-700 & 0.99 & & \\
\cline{1-11}

253-1/left      & 175 & MeOH & $578 \pm 28$  & \multirow{6}{*}{$568 \pm 11$} & $713 \pm 27$  & \multirow{6}{*}{$701 \pm 11$} & 20-60 & $84 \pm 10$ & 500-700 & 0.79 & & \\
253-1/right     & 175 & MeOH & $578 \pm 28$  &                               & $713 \pm 25$  &                               & 20-60 & $88 \pm 10$ & 500-700 & 0.79 & & \\
253-2/left      & 175 & MeOH & $567 \pm 28$  &                               & $695 \pm 30$  &                               & 20-60 & $82 \pm 9$  & 500-700 & 0.79 & & \\
253-2/right     & 175 & MeOH & $567 \pm 28$  &                               & $699 \pm 25$  &                               & 20-60 & $85 \pm 9$  & 500-700 & 0.78 & & \\
253-3/left      & 175 & MeOH & $559 \pm 25$  &                               & $695 \pm 24$  &                               & 20-60 & $81 \pm 9$  & 500-700 & 0.78 & & \\
253-3/right     & 175 & MeOH & $559 \pm 25$  &                               & $694 \pm 24$  &                               & 20-60 & $85 \pm 9$  & 500-700 & 0.78 & & \\
\hline

255-1/left      & 200 & -    & $1242 \pm 44$ & \multirow{6}{*}{$1248 \pm 19$} & $1283 \pm 35$ & \multirow{6}{*}{$1297 \pm 15$} & 20-60 & $85 \pm 9$  & 500-700 & 0.97 & \multirow{12}{*}{0.413 } & \multirow{12}{*}{0.471} \\
255-1/right     & 200 & -    & $1242 \pm 44$ &                                & $1294 \pm 44$ &                                & 20-60 & $87 \pm 9$  & 500-700 & 0.96 & & \\
255-2/left      & 200 & -    & $1253 \pm 49$ &                                & $1307 \pm 29$ &                                & 20-60 & $85 \pm 9$  & 500-700 & 0.96 & & \\
255-2/right     & 200 & -    & $1253 \pm 49$ &                                & $1306 \pm 35$ &                                & 20-60 & $89 \pm 9$  & 500-700 & 0.96 & & \\
255-3/left      & 200 & -    & $1248 \pm 44$ &                                & $1295 \pm 41$ &                                & 20-60 & $84 \pm 9$  & 500-700 & 0.96 & & \\
255-3/right     & 200 & -    & $1248 \pm 44$ &                                & $1300 \pm 37$ &                                & 20-60 & $88 \pm 9$  & 500-700 & 0.96 & & \\
\cline{1-11}

254-1/left      & 200 & MeOH & $567 \pm 29$  & \multirow{6}{*}{$565 \pm 13$} & $663 \pm 27$  & \multirow{6}{*}{$656 \pm 10$} & 20-60  & $85 \pm 9$ & 500-700 & 0.84 & & \\
254-1/right     & 200 & MeOH & $567 \pm 29$  &                               & $657 \pm 28$  &                               & 20-60  & $89 \pm 9$ & 500-700 & 0.84 & & \\
254-2/left      & 200 & MeOH & $575 \pm 37$  &                               & $660 \pm 23$  &                               & 20-60  & $84 \pm 9$ & 500-700 & 0.85 & & \\
254-2/right     & 200 & MeOH & $575 \pm 37$  &                               & $668 \pm 23$  &                               & 20-60  & $89 \pm 9$ & 500-700 & 0.84 & & \\
254-3/left      & 200 & MeOH & $552 \pm 27$  &                               & $640 \pm 24$  &                               & 20-60  & $83 \pm 9$ & 400-600 & 0.84 & & \\
254-3/right     & 200 & MeOH & $552 \pm 27$  &                               & $650 \pm 26$  &                               & 20-60  & $86 \pm 9$ & 400-600 & 0.83 & & \\
\hline

\end{tabular}%

\end{adjustbox}

\begin{tablenotes}[flushleft]\footnotesize
\item[$^a$] $^a$ Six signal intensity values were obtained per sample, because of three repeat line-scans, and since the structures were mirrored, the left and right sides of the central open area (Zone I) gave saturation profiles that were analyzed separately.
\item[$^b$] $^b$ Data shown as mean $\pm$ propagated standard deviation $\sigma$ (for $N = 6$). Propagated error was calculated as $\frac{1}{N}\sqrt{\sum \sigma_i^2}$.
\item $^c$Ratios were calculated using background-subtracted intensities: $\frac{\text{Zone I} - \text{Background}}{\text{Zone IIb} - \text{Background}}$.
\item[$^d$] $^d$ Ratios were taken of the signal corresponding to the process with and without the inhibitor. Ratios were calculated using background-subtracted intensities: $\frac{\text{With inhibitor} - \text{Avg. background}}{\text{Without inhibitor} - \text{Avg. background}}$.
\end{tablenotes}
\label{tab:signal_intensity}
\end{sidewaystable}

\begin{sidewaystable}[p]
\captionsetup[sidewaystable]{position=top}
\centering
\caption{SEM-EDS line scans: reaction conditions, penetration depths, and derived sticking coefficients. Average sticking coefficients are shown with standard deviations.$^a$}
\footnotesize
\setlength{\tabcolsep}{3pt}
\renewcommand{\arraystretch}{1.12}

\begin{adjustbox}{max height=0.97\textheight,keepaspectratio}
  \begin{tabular}{|M{1.6cm}|M{1.0cm}|M{3.0cm}|M{1.3cm}|M{1.2cm}|M{1.8cm}|M{1.6cm}|M{3.0cm}|M{1.9cm}|M{2.4cm}|}

\hhline{|t:==========:|}
\makecell[l]{\rule{0pt}{3.6ex}Sample\\name} &
\makecell[l]{\rule{0pt}{3.6ex}Code\\on chip} &
\makecell[l]{\rule{0pt}{3.6ex}SEM-EDS line scan\\number\\(full sample name)} &
\makecell[l]{\rule{0pt}{3.6ex}ALD\\ temp.\\(\si{\degreeCelsius})} &
\makecell[l]{\rule{0pt}{3.6ex}Inhibitor} &
\makecell[l]{\rule{0pt}{3.6ex}Penetration\\depth $x$\\(\si{\micro\meter})} &
\makecell[l]{\rule{0pt}{3.6ex}Penetration\\depth $x/H$\\(-)} &
\makecell[l]{\rule{0pt}{3.6ex}Slope of normalized\\intensity vs.\\dimensionless\\ distance $x/H$ (-)} &
\makecell[l]{\rule{0pt}{3.6ex}Sticking\\ coefficient\\(Arts et al.\cite{arts_sticking_2019})} &
\makecell[l]{\rule{0pt}{3.6ex}Average\\ sticking\\coefficient} \\
\hhline{|==========|}

AP-251 & D4 & 251-1/left  & 150 & -    & 151.9 & 303.9 & 0.0077 & 0.000817 & \makecell[l]{\scriptsize $8.60\times10^{-4}$\\\scriptsize $\pm\,1.39\times10^{-4}$} \\
AP-251 & D4 & 251-1/right & 150 & -    & 145.7 & 291.5 & 0.0078 & 0.000855 &  \\
AP-251 & D4 & 251-2/left  & 150 & -    & 149.5 & 299.0 & 0.0089 & 0.001089 &  \\
AP-251 & D4 & 251-2/right & 150 & -    & 148.3 & 296.6 & 0.0080 & 0.000885 &  \\
AP-251 & D4 & 251-3/left  & 150 & -    & 147.2 & 294.4 & 0.0079 & 0.000857 &  \\
AP-251 & D4 & 251-3/right & 150 & -    & 148.9 & 297.8 & 0.0069 & 0.000656 &  \\
\hhline{|----------|}

AP-252 & D3 & 252-1/left      & 150 & MeOH & 214.9 & 429.8 & 0.0053 & 0.000389 & \makecell[l]{\scriptsize $4.00\times10^{-4}$\\\scriptsize $\pm\,8.06\times10^{-5}$} \\
AP-252 & D3 & 252-1/right     & 150 & MeOH & 213.5 & 427.0 & 0.0050 & 0.000348 &  \\
AP-252 & D3 & 252-2/left      & 150 & MeOH & 217.4 & 434.8 & 0.0055 & 0.000417 &  \\
AP-252 & D3 & 252-2/right     & 150 & MeOH & 213.1 & 426.2 & 0.0056 & 0.000442 &  \\
AP-252 & D3 & 252-3/left      & 150 & MeOH & 215.4 & 430.9 & 0.0045 & 0.000285 &  \\
AP-252 & D3 & 252-3/right     & 150 & MeOH & 214.8 & 429.6 & 0.0061 & 0.000520 &  \\
\hhline{|----------|}

AP-256 & D6 & 256-1/left      & 175 & -    & 150.4 & 300.7 & 0.0191 & 0.005095 & \makecell[l]{\scriptsize $5.10\times10^{-3}$\\\scriptsize $\pm\,3.12\times10^{-4}$} \\
AP-256 & D6 & 256-1/right     & 175 & -    & 147.3 & 294.7 & 0.0197 & 0.005373 &  \\
AP-256 & D6 & 256-2/left      & 175 & -    & 148.6 & 297.2 & 0.0191 & 0.005095 &  \\
AP-256 & D6 & 256-2/right     & 175 & -    & 148.6 & 297.2 & 0.0189 & 0.004986 &  \\
AP-256 & D6 & 256-3/left      & 175 & -    & 149.2 & 298.3 & 0.0199 & 0.005480 &  \\
AP-256 & D6 & 256-3/right     & 175 & -    & 148.5 & 296.9 & 0.0182 & 0.004594 &  \\
\hhline{|----------|}

AP-253 & D5 & 253-1/left      & 175 & MeOH & 215.0 & 430.1 & 0.0076 & 0.000804 & \makecell[l]{\scriptsize $8.11\times10^{-4}$\\\scriptsize $\pm\,1.09\times10^{-4}$} \\
AP-253 & D5 & 253-1/right     & 175 & MeOH & 213.2 & 426.5 & 0.0080 & 0.000899 &  \\
AP-253 & D5 & 253-2/left      & 175 & MeOH & 213.8 & 427.6 & 0.0066 & 0.000605 &  \\
AP-253 & D5 & 253-2/right     & 175 & MeOH & 213.5 & 426.9 & 0.0077 & 0.000822 &  \\
AP-253 & D5 & 253-3/left      & 175 & MeOH & 214.8 & 429.7 & 0.0080 & 0.000900 &  \\
AP-253 & D5 & 253-3/right     & 175 & MeOH & 214.4 & 428.7 & 0.0078 & 0.000838 &  \\
\hhline{|----------|}

AP-255 & D7 & 255-1/left      & 200 & -    & 144.2 & 288.3 & 0.0320 & 0.014225 & \makecell[l]{\scriptsize $1.23\times10^{-2}$\\\scriptsize $\pm\,1.40\times10^{-3}$} \\
AP-255 & D7 & 255-1/right     & 200 & -    & 143.0 & 286.0 & 0.0293 & 0.011945 &  \\
AP-255 & D7 & 255-2/left      & 200 & -    & 143.4 & 286.8 & 0.0311 & 0.013440 &  \\
AP-255 & D7 & 255-2/right     & 200 & -    & 143.6 & 287.3 & 0.0288 & 0.011557 &  \\
AP-255 & D7 & 255-3/left      & 200 & -    & 145.0 & 289.9 & 0.0299 & 0.012418 &  \\
AP-255 & D7 & 255-3/right     & 200 & -    & 144.1 & 288.2 & 0.0272 & 0.010261 &  \\
\hhline{|----------|}

AP-254 & G9 & 254-1/left      & 200 & MeOH & 227.6 & 455.2 & 0.0071 & 0.000704 & \makecell[l]{\scriptsize $6.73\times10^{-4}$\\\scriptsize $\pm\,2.32\times10^{-4}$} \\
AP-254 & G9 & 254-1/right     & 200 & MeOH & 229.5 & 458.9 & 0.0087 & 0.001042 &  \\
AP-254 & G9 & 254-2/left      & 200 & MeOH & 227.1 & 454.3 & 0.0077 & 0.000817 &  \\
AP-254 & G9 & 254-2/right     & 200 & MeOH & 227.0 & 454.0 & 0.0063 & 0.000545 &  \\
AP-254 & G9 & 254-3/left      & 200 & MeOH & 227.7 & 455.5 & 0.0062 & 0.000528 &  \\
AP-254 & G9 & 254-3/right     & 200 & MeOH & 222.3 & 444.7 & 0.0054 & 0.000402 &  \\
\hhline{|b:==========:|}

\end{tabular}%
\end{adjustbox}

\begin{tablenotes}[flushleft]\footnotesize
\item[$^a$] $^a$ Six penetration depth and slope values were obtained per sample, because three repeat line-scans per sample, and since the structures were mirrored, the left and right sides of the central open area (Zone I) gave saturation profiles that were analyzed separately.

\end{tablenotes}
\label{tab:sem-eds-sideways}
\end{sidewaystable}

\clearpage

\begin{figure}[H]
\clearpage
    \centering
    \includegraphics[width=0.4\linewidth]{Supplementary_info/esi_figures/stickingtemp_2026-09-03.jpg}
    \caption{Lumped sticking coefficients back-extracted from the slope of the adsorption front of the experimental ALD saturation profiles (values are from Table \ref{tab:slopes_stickingcoeff}).}
    \label{fig:growin_sticking}
\end{figure}

\section{Derivation of the inverse square root PD-GPC relationship from published analytical model}

One can derive the inverse square root relation between penetration depth (PD) in LHAR cavities and the GPC in ALD, discovered in this work, from a published  analytical model. 

The Gordon et al.\cite{gordon2003kinetic} model has been derived considering that the LHAR cavity acts as a "vacuum pump" with zero coverage of adsorbed species, and up to the penetration depth ($\lambda$ in the Gordon et al.\cite{gordon2003kinetic} notation, $x_{PD}$ here), there is full coverage. The impingement flux of molecules with a known partial pressure ($p$ in the Gordon et al.\cite{gordon2003kinetic} notation, $p_{A0}$ here) is considered, Knudsen diffusion conditions only are considered and flux at  the distance (PD) up to which film has grown is reduced by a known Clausing Factor. 

We start from Equation 13 of Gordon et al. \cite{gordon2003kinetic}: 
\begin{equation}
Pt = S \sqrt{2\pi m k T}\,\bigl[\,4a + \tfrac{3}{2} a^2\,\bigr],
\end{equation}
where $P$ (Pa) is the partial pressure of the reactant (here $p_{A0}$),  $t$ (s) is the exposure time, $S$ (\#/$\text{m}^{2}$)  is the saturation concentration of precursor molecules per unit area (here $q_0$), $m$ (kg) is the molecular mass, $k$ ($\text{J}\cdot\text{K}^{-1}$) is the Boltzmann constant (here $k_B$), $T$ (K) is the substrate temperature, and $a$ is Gordon et al.'s \cite{gordon2003kinetic} aspect ratio for the coating. This equation can be simplified considering only the second-order term with $a$, because that dominates for high aspect ratios e.g. over 100:1: 
\begin{equation}
Pt \simeq S \sqrt{2\pi m k T}\,\bigl[\tfrac{3}{2} a^2\,\bigr].
\label{eq:gordon_secondorder}
\end{equation}

To convert the simplified equation to the rectangular LHAR geometry and symbols used in this work, we interpret the "aspect ratio" of Equation 4 by Gordon et al. \cite{gordon2003kinetic} in the rectangular LHAR cavity geometry. The $a={Lp}/{4A}$, where $p$ is the perimeter of the structure (for rectangular LHAR $p=2(W+H)$) and $A$ is the cross-sectional area of the cavity (for rectangular LHAR $A=WH$). For a full LHAR structure length $L$, we have $a={2L(W+H)}/{4WH}$. Recognizing that in this work's LHAR structures $W \gg H$ ($W/H \to \infty$), the equation simplifies to $a = L / 2H$. Further,  replacing the $L$ (length of the channel) with the actual coating depth $x_{\mathrm{PD}}$ similarly as done by Gordon et al.\cite{gordon2003kinetic} for $\lambda$, we get 
\begin{equation}
 a = x_{\mathrm{PD}} / 2H  . 
\end{equation}

Using our notations for the the rectangular LHAR related $a = x_{\mathrm{PD}} / 2H$, $S = q_0$, $P = p_{A0}$ and $k = k_B$, Equation 13 of Gordon et al.\cite{gordon2003kinetic} after simplification (Equation \ref{eq:gordon_secondorder}) transforms to:
\begin{equation}
    p_{A0} t = q_0 \sqrt{2\pi m k_B T} \times \frac{3}{2} \left( \frac{x_{PD}}{2H} \right)^2
\end{equation}

\noindent Solving  for $x_{\mathrm{PD}}$, we get:



\begin{equation}
    x_{\mathrm{PD}} = \sqrt{\frac{8}{3} \frac{ p_{A0} t}{q_0 \sqrt{2\pi m k_B T}}} \cdot H.
\end{equation}
This is the sought-after inverse square root relationship.

In conclusion, an inverse square root relationship between penetration depth ($x_{PD}$) and growth per cycle ($q_0$), as found in this work both through simulations and experiments, can be inferred for Knudsen diffusion conditions from the analytical model by Gordon et al.\cite{gordon2003kinetic} Detailed analysis of the analytical equations by Ylilammi et al. \cite{ylilammi2018modeling} (not shown) would lead to the same observed dependency, which would even extend to  molecular diffusion (low Knudsen numbers). The simulations of this work (made with the full diffusion--reaction model, not with simplified analytical equations) support the existence of the observed inverse square root relation only down to ca. Kn $\sim 1$.
\clearpage

\section{List of symbols }
\begin{enumerate}[align=left, noitemsep, labelwidth=60pt, labelsep=1.5em, leftmargin=85pt]
\item[$a$] Average adsorption site area (containing a single payload atom M) on the substrate ($\mathrm{m^{2}}$)
\item[$\bar{a}_{\mathrm{MZ}_x}$] Theoretical average surface area occupied per metal (M) atom in an ideal monolayer of an ALD-grown film material (MZ$_x$) ($\mathrm{m^{2}}$)
\item[$b$] Exponent in the power-law fitting equation $(x/H)_{\theta=0.5} = C q_0^b$ relating penetration depth to the adsorption capacity (-)
\item[$c$] (Lumped) sticking coefficient (-)
\item[$C$] Proportionality constant in the power-law fitting equation $(x/H)_{\theta=0.5} = C q_0^b$ relating penetration depth to the adsorption capacity (-)
\item[$D$] Characteristic feature dimension (m)
\item[$D_\mathrm{A}$] Molecular diffusion coefficient ($\mathrm{m^2\,s^{-1}}$)
\item[$D_\mathrm{eff}$] Effective diffusion coefficient ($\mathrm{m^2\,s^{-1}}$)
 \item[$D_\mathrm{Kn}$] Knudsen diffusion coefficient ($\mathrm{m^2\,s^{-1}}$)

 \item[$d_\mathrm{A}$] Hard-sphere diameter of molecule A (m)
\item[$d_\mathrm{I}$] Hard-sphere diameter of the inert gas
molecule (m)
\item[$g$] Net adsorption rate ($\mathrm{m^{-2}\ s^{-1}}$)
\item[$H$] Height of the channel (m)
\item[$h$] Hydraulic diameter of the channel ($\mathrm{m}$)
\item[$\bar{h}_{\mathrm{ml}}$] Monolayer thickness of the ALD film material MZ$_x$ ($\mathrm{m}$)

\item[$\mathrm{Kn}$] Knudsen number (-)
\item[$k_{\mathrm{B}}$] Boltzmann constant (J\,K$^{-1}$)
\item[$L$] Length of the channel (m)
\item[$\lambda$] Mean free path (m)
\item[$M$] Molar mass of one formula unit of the ALD-grown film material ($\mathrm{MZ}_x$) ($\mathrm{kg\,mol^{-1}}$)
\item[$M_\mathrm{A}$] Molar mass of the reactant A ($\mathrm{kg\, mol^{-1}}$)
\item[$M_\mathrm{I}$] Molar mass of the inert gas I ($\mathrm{kg\, mol^{-1}}$)
\item[$m_\mathrm{A}$] Molecular mass of reactant A (kg)
\item[$m_\mathrm{I}$] Molecular mass of inert gas I (kg)

\item[$N_\mathrm{0}$]Avogadro’s constant ($\mathrm{mol^{-1}}$)
\item[$P_\mathrm{d}$] Desorption probability  ($\mathrm{s^{-1}}$)
\item[$p_{\mathrm{A}}$] Partial pressure of reactant A ($\mathrm{Pa}$)
\item[$p_{\mathrm{A0}}$] Initial partial pressure of reactant A at the beginning of the channel ($\mathrm{Pa}$)
\item[$p_{\mathrm{I}}$] Partial pressure of inert gas I ($\mathrm{Pa}$)

\item[$Q$] Collision rate of reactant A with the surface at unit pressure in the Ylilammi \textit{et al.}\cite{ylilammi2018modeling} model ($\mathrm{m^{-2}\ s^{-1}\ Pa^{-1}}$)

\item[$q_0$] Adsorption capacity of metal atoms (M) per unit area in the ALD
growth of film of the $\mathrm{MZ_x}$ material (\#/$\mathrm{m^{2}}$)

\item[$\bar{q}_\mathrm{ml}$] Surface density of metal atoms (M) per unit area in an average monolayer of ALD
grown $\mathrm{MZ_x}$ material (\#/$\mathrm{m^{2}}$)

\item[$Q$] Collision rate of reactant A with the surface at unit pressure\cite{ylilammi2018modeling}  ($\text{m}^{-2}\ \text{s}^{-1}\ \text{Pa}^{-1}$)
\item[$\rho$] Mass density of the ALD film material ($\mathrm{kg\,m^{-3}}$)
\item[$R$] Gas constant ($\mathrm{J\ K^{-1}\ mol^{-1}}$)
\item[$S$]   Saturation concentration of precursor molecules per unit area in the Gordon et al. model\cite{gordon2003kinetic} ($\#/\mathrm{m^{2}}$), (equivalent to the adsorption capacity $q_0$ in this work).
\item[$\sigma_{i,j}$] Collision cross-section between molecules $i$ and $j$ ($\mathrm{m^{2}}$)
\item [${t}$] Time (s)
\item [${T}$] Temperature (K)
\item [${\theta}$] Surface coverage (-)
\item [$\overline{v}_A$] Thermal velocity of molecule A ($\mathrm{m~s^{-1}}$)
\item[$W$] Width of the channel (m)
\item[$x$] Physical distance (m)
\item[$x_{\mathrm{PD}}$] Physical film penetration depth (or coating depth) in the channel ($\mathrm{m}$)
\item[$x/H$] Dimensionless distance, i.e. the ratio of the
physical distance to the channel height (-)
\item[$x/L$] Normalized distance, i.e. the ratio of the
physical distance to the channel length (-)
\item[$(x/H)_{\theta=0.5}$]  Film penetration depth at half surface coverage in terms of dimensionless distance, representing the distance at which the film surface coverage (and hence thickness) reduces to 0.5 (-)
\item[$(x/H)_{\theta=0.42}$] Film penetration depth at surface coverage of 0.42 in terms of dimensionless distance, representing the distance at which the film surface coverage reduces to 0.42 (-)
\item[$\tilde{z}_{\mathrm{A}}$] Collision frequency of reactant A with other gas molecules in a mixture of reactant A and inert gas I ($\mathrm{s^{-1}}$)
\end{enumerate}
\clearpage

\bibliography{Maindoc/references} 
\bibliographystyle{rsc} 